\documentclass[journal]{IEEEtran}
\usepackage{algorithm,amsbsy,amsmath,amssymb,epsfig,bbm,mathrsfs,multirow,amsthm}
\usepackage{array,multirow,graphicx}
\usepackage{graphicx}
\usepackage{graphics}
\usepackage{subcaption}
\usepackage{epstopdf}
\usepackage{color}
\usepackage{bbm}
\usepackage{comment}
\usepackage{cite}
\usepackage{algpseudocode}
\usepackage{soul}

\renewcommand{\baselinestretch}{1.0}

\newcommand{\beq}{\begin{equation}}
\newcommand{\eeq}{\end{equation}}
\newcommand{\bitm}{\begin{itemize}}
\newcommand{\ba}{\begin{array}}
\newcommand{\ea}{\end{array}}
\newcommand{\eitm}{\end{itemize}}
\newcommand{\beqn}{\begin{eqnarray}}
\newcommand{\eeqn}{\end{eqnarray}}
\newcommand{\beqno}{\begin{eqnarray*}}
\newcommand{\eeqno}{\end{eqnarray*}}
\newcommand{\bma}{\begin{displaymath}}
\newcommand{\ema}{\end{displaymath}}
\newcommand{\bnu}{\begin{enumerate}}
\newcommand{\enu}{\end{enumerate}}
\newcommand{\bce}{\begin{center}}
\newcommand{\ece}{\end{center}}
\newcommand{\btb}{\begin{tabular}}
\newcommand{\etb}{\end{tabular}}

\begin{document}
\title{\huge Joint Time Scheduling and Transaction Fee Selection in Blockchain-based RF-Powered Backscatter Cognitive Radio Network}

\author{ 
\IEEEauthorblockN{Tran The Anh,
Nguyen Cong Luong, Zehui Xiong, Dusit Niyato,~\IEEEmembership{Fellow,~IEEE}, Dong In Kim,~\IEEEmembership{Fellow,~IEEE}}

\thanks{T. T. Anh, Z.~Xiong and D. Niyato are with the School of Computer Science and Engineering, Nanyang Technological University, Singapore. Emails: theanh.tran@ntu.edu.sg, ZXIONG002@e.ntu.edu.sg, dniyato@ntu.edu.sg.}
\thanks{N. C.~Luong is with Faculty of Information Technology, PHENIKAA University, Hanoi 12116, Vietnam, and is with PHENIKAA Research and Technology Institute (PRATI), A\&A Green Phoenix Group JSC, No.167 Hoang Ngan, Trung Hoa, Cau Giay, Hanoi 11313, Vietnam. Email: luong.nguyencong@phenikaa-uni.edu.vn.}
\thanks{D.~I.~Kim is with School of Information and Communication Engineering, Sungkyunkwan University, Korea. Email: dikim@skku.ac.kr.}
\vspace{-0.7cm}
}
\maketitle
\begin{abstract}
In this paper, we develop a new framework called \textit{blockchain-based Radio Frequency (RF)-powered backscatter cognitive radio network}. In the framework, IoT devices as secondary transmitters transmit their sensing data to a secondary gateway by using the RF-powered backscatter cognitive radio technology. The data collected at the gateway is then sent to a blockchain network for further verification, storage and processing. As such, the framework enables the IoT system to simultaneously optimize the spectrum usage and maximize the energy efficiency. Moreover, the framework ensures that the data collected from the IoT devices is verified, stored and processed in a decentralized but in a trusted manner. To achieve the goal, we formulate a stochastic optimization problem for the gateway under the dynamics of the primary channel, the uncertainty of the IoT devices, and the unpredictability of the blockchain environment. In the problem, the gateway jointly decides (i) the time scheduling, i.e., the energy harvesting time, backscatter time, and transmission time, among the IoT devices, (ii) the blockchain network, and (iii) the transaction fee rate to maximize the network throughput while minimizing the cost. To solve the stochastic optimization problem, we then propose to employ, evaluate, and assess the Deep Reinforcement Learning (DRL) with Dueling Double Deep Q-Networks (D3QN) to derive the optimal policy for the gateway. The simulation results clearly show that the proposed solution outperforms the conventional baseline approaches such as the conventional Q-Learning algorithm and non-learning algorithms in terms of network throughput and convergence speed. Furthermore, the proposed solution guarantees that the data is stored in the blockchain network at a reasonable cost.
\end{abstract}
\begin{IEEEkeywords}
Cognitive radio, ambient backscatter,
RF energy harvesting, blockchain, time scheduling, deep reinforcement learning, IoT.
\end{IEEEkeywords}

\section{Introduction}

Internet of Things (IoT) systems require the massive deployment of the communication devices, thereby facing a big issue of scarcity of spectrum resource. To address both the radio spectrum shortage and the energy constraint in the IoT systems, Radio Frequency (RF)-powered backscatter cognitive radio \cite{huynh2018}, \cite{li2018}, \cite{kang2018}, \cite{liu2013} has been recently used as a promising solution. With the RF-powered backscatter cognitive radio technology, IoT devices as secondary transmitters are able to harvest energy from primary signals, e.g., broadcast from the primary transmitter, and then use the harvested energy to actively transmit their data to a secondary gateway. Moreover, the IoT devices are able to backscatter their data to the secondary gateway by modulating and reflecting the surrounding ambient RF signals. Therefore, the RF-powered backscatter cognitive radio is an effective solution to improve the network performance, i.e., the network throughput, in the IoT systems. 

However, the IoT systems face other serious issues. Specifically, the data collected from the IoT devices is still centralized in a server or in the cloud~\cite{khan2017cognitive}. This poses the transparency and traceability issues in which the data can be modified arbitrarily by unknown persons and applications. Moreover, the security issues arise since the central entity, i.e., the cloud, is vulnerable to cyber attacks. In addition, given a large number of IoT devices, the communication and bandwidth cost increases. Low efficiency, speed, and reliability due to the bottleneck and a single point of failure are also remaining critical issues.Thus, a new solution of the data management needs to be designed. 

Recently, blockchain technology \cite{schrijvers2016} has brought many promising potentials which has been applied in various applications including data management for IoT systems~\cite{khan2018}, \cite{kim2019blockchained}. In particular, blockchain is regarded as a decentralized database, i.e., a ledger~\cite{neisse2017}, in which transactions including IoT data are recorded and processed by a number of nodes in the whole blockchain network instead of a centralized authority or a single entity. The blockchain also enhances the security and guarantees the data integrity since the transactions must be agreed and verified by the nodes before they are recorded \cite{liu2017}. For this reason, blockchain can be integrated with the RF-powered backscatter cognitive radio, namely \textit{blockchain-based RF-powered backscatter cognitive radio} \cite{saghiri2018}. 

The blockchain-based RF-powered backscatter cognitive radio has three key benefits. First, it enables the IoT systems to improve the spectrum usage since the IoT devices, as secondary transmitters, can use primary channels, i.e., of the primary transmitter, for data transmissions when the channels are not occupied. Second, the blockchain-based RF-powered backscatter cognitive radio improves the energy efficiency since the IoT devices are able to harvest energy or backscatter data when the primary channels are occupied. Third, the data from the IoT devices sent to the blockchain is verified, recorded and processed in a decentralized and trusted manner~\cite{kim2019blockchained}. However, to achieve the goals with a reasonable storage cost, it is required to design an optimal mechanism for the gateway to decide the time scheduling among the IoT devices, blockchain network, and transaction fee rate. This is challenging due to the following reasons. First, the states, i.e., the busy and idle states, of the primary channel are dynamic. Second, the states including sensing data state and energy state of the IoT devices are uncertain. Third, the states of the blockchain networks are unpredictable. To address the challenge, we formulate a stochastic optimization problem for the gateway under the dynamics, uncertainty, and unpredictability of the system. The optimization problem allows the gateway to decide (i) the time scheduling, i.e., backscatter time, energy harvesting time, and active transmission time, among multiple IoT devices, (ii) blockchain network, and (iii) transaction fee rate to maximize the network throughput. The optimization problem accounts for the states of IoT devices, the primary channel, and the blockchain networks. The state spaces may be very large, and the optimization problem becomes very complex. Therefore, we propose to employ and assess an advanced learning algorithm, i.e., the DRL with Dueling Double Q-Networks (D3QN)~\cite{wang2016}, to solve the problem. As demonstrated in the simulation results, our proposed approach can not only effectively deal with the dynamics, uncertainty, and large state space of the system but also significantly improve the system performance compared with the existing time scheduling approaches. The main contributions of this paper are as follows:
\begin{itemize}
\item We propose a multi-user blockchain-based RF-powered backscatter cognitive radio network. This network enables the IoT systems to simultaneously optimize the spectrum usage and the energy efficiency to maximize their performance. Moreover, sensing data from the IoT devices is verified, recorded and processed in a decentralized and trusted manner. The term "trusted'' means that the sensing data of the IoT devices is not arbitrarily modified when the data is stored in the blockchain. 
\item We formulate a stochastic optimization problem for the blockchain-based RF-powered backscatter cognitive radio network. The optimization problem enables the gateway to jointly decide the time scheduling among the IoT devices, blockchain network, and transaction fee rate. The objective is to maximize the network throughput with the optimal time schedudling. The joint optimization also minimizes the storage cost with the appropriate blockchain selection and transaction fee rate decisions.
\item To jointly achieve the optimal time scheduling, blockchain network, and transaction fee rate decision policy for the gateway, we propose to employ, evaluate, and assess the DRL algorithm to solve the novel problem formulation under the system model. In particular, we use the D3QN that is able to (i) handle the bigger and more complex problem than traditional approaches, (ii) overcome the instability of the learning, and (ii) reduce the overestimation of action values. 
\item Finally, we perform extensive simulations to demonstrate the efficiency of our proposed approach in comparison with the conventional approaches such as the conventional Q-learning and non-learning algorithms. We reveal that the DRL with D3QN can achieve better performance in terms of network throughput, transaction fee, and convergence speed compared with the conventional baseline approaches. We further evaluate the proposed DRL scheme. The simulation results show that the performance of the DRL significantly improves as the number of blockchain networks
increases. 
\end{itemize}

The rest of this paper is organized as follows. Section  II presents a brief review of the related work. Section III describes the system model. Section IV presents the problem formulation. Section V introduces the DRL algorithm for the time scheduling, blockchain network selection, and transaction fee rate decisions in the blockchain-based RF-powered backscatter cognitive radio network. Section VI discusses the performance evaluation results. Section VII summarizes the paper.

\section{Related Works}
To improve the network performance and to address the energy constraint in the IoT systems, RF-powered backscatter cognitive radio has been recently used as a promising solution. With the RF-powered backscatter cognitive radio technology, IoT devices are able to harvest energy from primary signals and then use the harvested energy to actively transmit their data to a secondary gateway. Moreover, the IoT devices are able to backscatter their data to the secondary gateway by modulating and reflecting the surrounding ambient RF signals. However, how to optimize the  time scheduling and admission control in the IoT systems is a big issue.

The authors in~\cite{hoang2017} addressed the data scheduling and admission control problem for a gateway in a backscatter sensor network. The problem is to find the optimal data collection policy to minimize the weighted sum of delay of different types of data from sensors. The Markov decision process and the reinforcement learning algorithm based on the linear function approximation method are adopted to solve the problem. Different from~\cite{hoang2017}, the authors in ~\cite{lyu2018a} formulated an optimization problem for a secondary transmitter, i.e., the secondary user, in an ambient backscatter communications network. The optimization problem is to derive an optimal control policy for sleep and active mode switching and reflection coefficient in the active mode. To solve the optimization problem, a two-stage method is proposed to obtain the optimal solution. However, only one single secondary user is considered. The authors in~\cite{lyu2018b} extended the work in~\cite{lyu2018a} to a hybrid backscatter assisted cognitive wireless powered radio network with multiple secondary users. In particular, the hybrid Harvest-Then-Transmit (HTT) and backscatter communications are adopted and integrated for the secondary users. The authors then formulated the optimal time allocation between the ambient backscatter mode and energy harvesting and that between the bistatic scatter mode and the HTT mode to maximize the throughput of the secondary system. The Lagrange multipliers method with the Karush-Kuhn-Tucker (KKT) conditions is then adopted to solve the problem. 



To model the time scheduling in RF-powered backscatter cognitive radio networks, game theory such as Stackelberg game can also be used as proposed in \cite{wang2018}. In the game approach, the gateway is the leader, and the secondary transmitters, i.e., IoT devices, are the followers. The gateway first determines spectrum sensing time and an interference price to maximize its revenue. Based on the time and price, each secondary transmitter determines the energy harvesting time, backscattering time, and transmission time so as to maximize its throughput. However, the proposed game approach requires complete and perfect sensing probability information, and thus it cannot deal with the dynamics of the network environment. Apart from the Stackelberg game, auction theory can be used for the time scheduling in the RF-powered backscatter cognitive radio network as proposed in \cite{gao2019}. In the auction approach, secondary transmitters, i.e., IoT devices, are buyers, and the gateway is the seller, i.e., the auctioneer. Each secondary transmitter submits its bid to the gateway to compete for the time resource. The bid includes the transmission demand and the valuation, i.e., the valuation of transmitting unit data to the secondary transmitter. After collecting the bids, the gateway performs the winner determination and time scheduling and determines the prices that the winning secondary transmitters need to pay. The heuristic algorithm with the generalized Vickrey-Clarke-Groves (VCG) pricing scheme are then adopted to solve the problem of the gateway. In fact, the cognitive radio network environment may be dynamic and uncertain. Therefore, the authors in \cite{anh2019} proposed to employ the Deep Reinforcement Learning (DRL)~\cite{mnih2015},~\cite{luong2019surveydrl} that enables the gateway to learn and derive the optimal time scheduling polity to maximize the network throughput. 

Although the RF-powered backscatter cognitive radio is an effective solution to improve the network throughput of the IoT systems, the data collected from the IoT devices is still centralized in a server or in the cloud. The centralized data aggregation scheme faces serious issues such as the transparency, traceability, privacy and security. Blockchain has been recently adopted for the data management in the IoT systems \cite{kotobi2017}, \cite{raju2017}, \cite{nguyen2019}. In the form of a chain of blocks, blockchain is actually a tamper-proof, distributed database or
ledger that records transactional data in a decentralized
Peer-to-Peer (P2P) network. As such, blockchain enables transactions including sensing data to be recorded and processed by a number of nodes over the whole network instead of a centralized authority. Since the transactions must be agreed and verified by the nodes before being recorded, blockchain enhances the security and guarantees the data integrity. In this regard, blockchain can be adopted to support an IoT crowdsensing market as proposed in \cite{shaohan2019}. The market consists of multiple sensing clouds, i.e., sellers, and multiple data users, i.e., buyers. The sensing clouds perform data sensing and transmit their data to the blockchain for distributed ledger services such as data/transaction verification and trading. The data users buy sensing service from the sensing clouds for their own tasks, i.e., data analytics. To access the sensing service, the data users rent a fraction of sensors in each sensing cloud by responding to a smart contract of the sensing clouds over the blockchain. As such, the service access and data delivery processes are completely self-organized in the form of smart contract without the requirement of an intermediary. From the perspective of the sensing clouds and the data users, the blockchain can be thus regarded as a decentralized platform as a service. The sensing clouds and the users are self-interested, and a hierarchical differential game is then used to model the interaction between them. The equilibrium of the hierarchical differential game is finally analyzed based on the Cauchy-Lipschitz theorem~\cite{Cauchytheorem}.


\section{System Model}
\begin{figure*}[t!]
 \centering
\includegraphics[width =15.5cm, height=10.3cm]{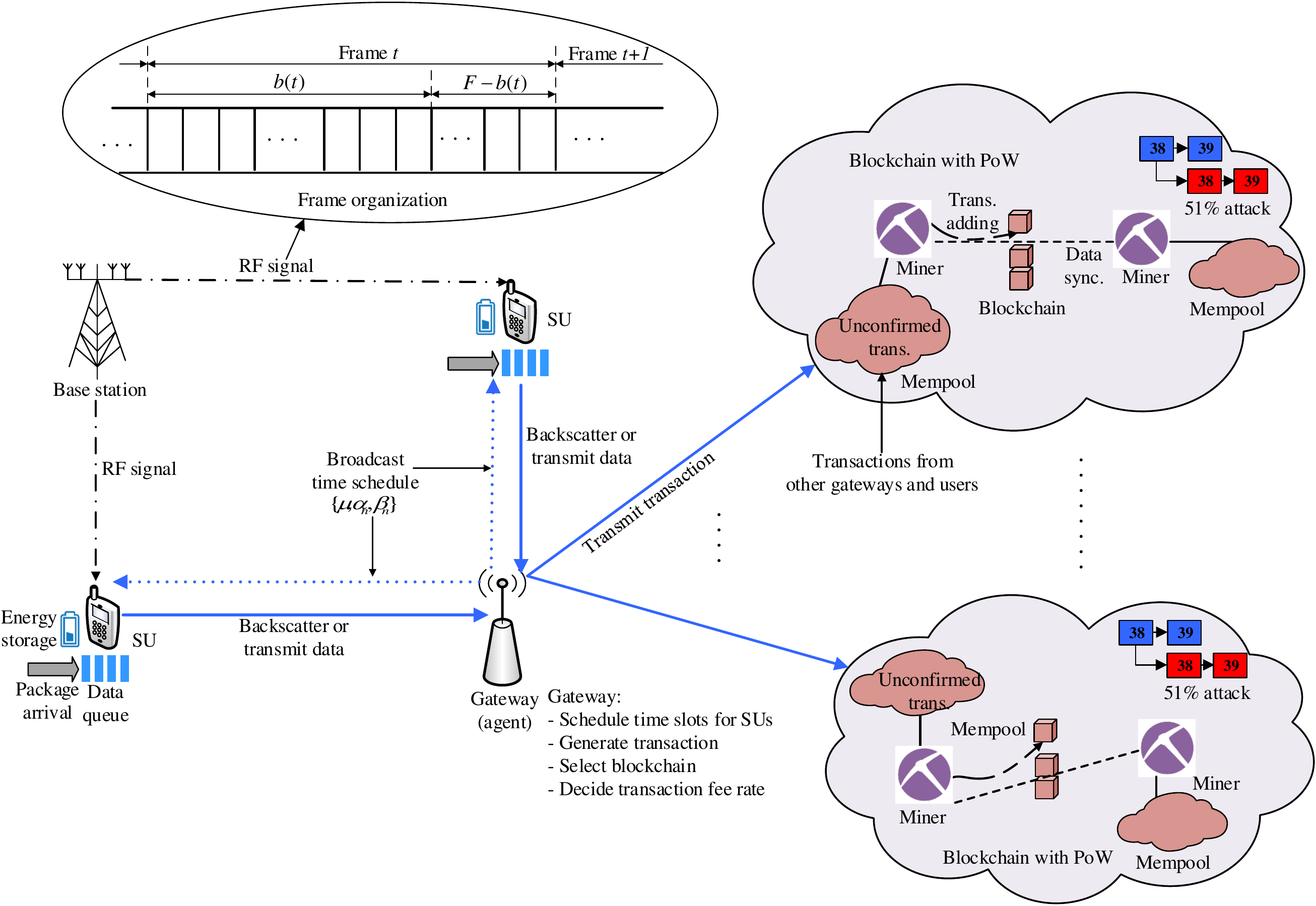}
 \caption{Blockchain-based RF-powered backscatter cognitive radio network.}
  \label{system_model}
\end{figure*}
We consider a blockchain-based RF-powered backscatter cognitive radio network as shown in Fig.~\ref{system_model}. The network consists of the primary transmitter, $N$ IoT devices as secondary transmitters, a secondary gateway, and $K$ blockchain networks. Each secondary transmitter is able to use the primary signal from the primary transmitter to harvest energy and backscatter data. Also, it can actively transmit data to the gateway when the primary channel is not occupied. To coordinate the data transmissions among the secondary transmitters, the gateway as a network controller decides the time schedule for the secondary transmitters. The gateway then broadcasts the time schedule to the network. Based on the time schedule, each secondary transmitter can harvest energy, backscatter data, or actively transmit data to the gateway. The gateway selects one of blockchain networks and determines the corresponding transaction fee rate~\cite{weber2017availability}. Note that we here consider a public or consortium blockchain in which the block size is limited, i.e., up to $1$ MB, and there is a number of transactions from other gateways and blockchain users. The gateway needs to thus decide the appropriate transaction fee rate such that its transaction will be early added to the block with a reasoable storage cost. The gateway transmits the data to the selected blockchain network as blockchain transactions for further verification, storage and processing.

As such, the blockchain-based RF-powered backscatter cognitive radio network is composed of two parts, i.e., the RF-powered backscatter cognitive radio network and the blockchain mining pool.


The primary transmitter transmits RF signal on a license channel. The transmission is organized into frames, and each frame is composed of $Y$ time slots. In Frame $t$ (see Fig.~\ref{system_model}), the primary transmitter uses $b(t)$ time slots for its own transmission. $b(t)$ is called busy channel period that is assumed to be random. The secondary transmitters transmit their data to the mining pool through the gateway. The gateway also schedules time slots for the secondary transmitters. In particular, during the busy channel period, the gateway assigns the time slots to all the secondary transmitters for energy harvesting. The number of time slots for energy harvesting is denoted by $\mu(t)$. Let $e^{\mathrm{h}}_n$ denote the number of energy units that secondary transmitter $n$ harvests in one busy time slot. The harvested energy is stored in energy storage, e.g., a super-capacitor, of the secondary transmitter. The capacity of the energy storage is denoted by $C_n$. The secondary transmitter has a data queue which stores an incoming packet, e.g., from its sensor device. Let $Q_n$ denote the maximum capacity of the data queue and $\gamma_n$ denote the probability that a packet arrives at the queue. The gateway assigns $b(t)-\mu(t)$ time slots of the busy channel period to the secondary transmitters for transmitting their data by using backscatter communications. This is called \textit{backscatter mode}. The number of time slots assigned to secondary transmitter $n$ is denoted by $\alpha_n(t)$, and the number of data units transmitted in each time slot in the backscatter mode is $d^{\mathrm{b}}_n$. During the idle channel period that has $Y-b(t)$ time slots, the secondary transmitters transmit their data to the gateway by using the license channel. This is called \textit{active mode}. The number of time slots assigned to secondary transmitter $n$ for the data transmission in the active mode is denoted by $\beta_n(t)$. In each time slot of the active mode, the secondary transmitter transmits $d^{\mathrm{a}}_n$ data units from the data queue, and the secondary transmitter consumes $e^{\mathrm{a}}_n$ energy units from its energy storage. Note that there may be data transmission errors due to the low channel quality. The data transmission in the backscatter mode and in the active mode is successful with probabilities $S^{\mathrm{b}}_n$ and $S^{\mathrm{a}}_n$, respectively. 
\subsection{Blockchain mining pool}
Upon receiving data from the secondary transmitters, the gateway creates the packet. Note that the size of the data packet can be different for each time frame depending on the number of data units that the gateway receives during the time frame. The gateway then observes data states, energy state, and blockchain networks' states that are described and presented in the next section. Based on the observations, the gateway selects one of the blockchain networks for storing the data. The gateway then creates a transaction that includes the data packet and determines a transaction fee rate denoted by $F$. Here, the transaction fee rate $F$ is the cost that the gateway is willing to pay the selected blockchain for storing one data unit. The gateway then transmits the transaction to the blockchain network for further processing. Here, the public or consortium blockchain which operates with the Proof-of-Work (PoW) protocol~\cite{kumar2019proof} is used to guarantee the high integrity for the IoT data. Note, however, that the blockchain based on Proof-of-Stake (PoS) and other consensus protocols can also be adopted straightforwardly in the system model. An integral of the blockchain is a consensus process in which a number of consensus nodes distributed across the network are required to complete the transaction validation and the block mining task. The consensus nodes are also called \textit{miners}. Each miner, say miner $i$, has a memory pool or ``mempool'', i.e., a waiting area, to store transactions sent from the gateway. In our work, we assume that the mempools of the miners have the same capacity that is $M_{\max}$ data units. Note that at a certain time slot, several transactions, e.g., transmitted from other gateways and network users, may arrive at the blockchain, and they are stored in the mempool of the miner. The transactions stored in the mempool are called \textit{unconfirmed transactions} that are broadcast and stored in mempools of all the miners in the network. Then, the consensus process works as follows. Each miner first collects transactions from its mempool and aggregates a set of  transaction records into a block. Then, the miners find a nonce value and add it into the block such that the hash value of the block is below a preset threshold~\cite{yang2018blockchain}. The process is known as \textit{mining}. Once the nonce value is found, the corresponding miner broadcasts its successfully mined block to the whole network for verification. If the majority of the miners agree that all transactions in the new block are valid, the block is linked to the main blockchain and the miner obtains a reward that consists of a fixed bonus and the transaction fee. In general, the probability that the transaction of the gateway is successfully added to the block at the current time slot is higher if the transaction is assigned with a higher fee rate~\cite{transaction_fee}. The reason is that the miners may spend substantial computing power and energy when verifying a block of transactions from the mempool. Thus, the miners need to be incentivized for the resources that they use to verify the unconfirmed transactions. Apart from the fee rate, the probability that the transaction of the gateway is successfully added to the block depends on the maximum size of the block, denoted by $B_{\max}$. In particular, as $B_{\max}$ is small, the block is not able to include a large number of transactions and the probability becomes lower.

As the block is linked to the main blockchain, transactions included in the block cannot be altered. Therefore, blockchain guarantees the security and the trustworthiness of IoT data stored in the blockchain. However, the blockchain system is vulnerable to a potential attack called \textit{51\% attack}\footnote{Our model is flexible, and other attacks, e.g., Sybil attack, can be considered. However, to make the model understandable and tractable, we consider the 51\% attack attack which is the most primitive attack to any blockchain networks.} or \textit{double-spending attack} \cite{karame2012double}. Such an attack may happen as one miner or a group of miners manages more than 50\% of the network hashing power, i.e., computing power. By controlling the majority of the computing power on the network, the attacker can interfere with the process of recording new blocks of other miners, i.e., honest miners, or maliciously modify the transactions. Therefore, the transaction transmitted from the gateway can be attacked. Let $p_a$ denote the probability that the transaction is attacked. As analyzed in~\cite{rosenfeld2014analysis}, $p_a$ depends on the numbers of blocks found by the honest miners and the attacker in the network. $p_a$ is defined as follows:
\begin{equation}
\label{double_spending_attack}
p_a = \left\{\begin{matrix}
1-\sum_{m=0}^{n} \begin{pmatrix}
m+n-1 \\ m
\end{pmatrix} 
\left ( p^n q^m-p^m q^n \right ) & \text{if  } q<p,\\
1 & \text{if  } q \geq p,
\end{matrix}\right.
\end{equation}
where $n$ and $m$ are, respectively, the numbers of blocks that are found by the honest network, i.e., the network of honest miners, and the attacker. $p$ and $q$, $p+q=1$, are the probabilities that a block is found by the honest network and the attacker, respectively. It is worth discussing the relationship between $q$ and the transaction fee rates. Typically, a transaction that has a high data value corresponds to a higher transaction fee rate. The attacker thus prefers to launch the attack to the block including such a transaction since this is more effective. The probability $q$ that a block is found by the attacker is high if the transactions included in the block have high data values. 

As such, even if the transaction is added to the new block, the transaction can still be attacked by the 51\% attack with probability $p_a$ as determined in~(\ref{double_spending_attack}). The transaction transmission is considered to be ``successful'' if the transaction is added to the new block at the current time slot and is not attacked. The objective of the gateway is to maximize the number of transactions and data units successfully stored in the blockchain system by jointly finding an optimal time scheduling among the secondary transmitters and determining the optimal values of the transaction fee rate. 


\section{Problem Formulation}
\label{sec:problem_formulation}
The state of the primary channel is dynamic, and the states, i.e., the energy and data queue, of the secondary transmitters are uncertain. Also, the blockchain environments' states including the mempool states are unpredictable. To achieve the objectives of the gateway, we thus formulate a stochastic optimization problem. The problem is defined by a tuple $<{\mathcal{S}}, {\mathcal{A}}, {\mathcal{P}}, {\mathcal{R}}>$, where ${\mathcal{S}}$ is the state space of the network or the gateway, ${\mathcal{A}}$ is the action space, ${\mathcal{P}}$ is the state transition probability function with $P{(s'|s,a)}$ being the probability that the current state $s \in  {\mathcal{S}}$ transits to the next state $s' \in {\mathcal{S}}$ when action $a \in {\mathcal{A}}$ is taken, and ${\mathcal{R}}$ is the reward function of the network or the gateway.

\subsection{State space}
Observing the network environment helps the gateway to learn and find its optimal policy in a faster and more effective manner. In particular, the gateway needs to  decide the energy harvesting time, the data backscatter time, and the active data transmission time, among the secondary transmitters to maximize the network throughput. Therefore, the network states should include states of the primary channel and those of the secondary transmitters. Furthermore, the gateway must decide an appropriate blockchain network for storage and a transaction fee rate such that the transaction is successfully added to the block with the minimum storage cost. The network state also includes the states of the blockchain environment. 

Let $\mathcal{S}^c$ denote the channel state. $\mathcal{S}^c$ represents the number of busy time slots and is defined as follows:
\begin{equation}
	{\mathcal{S}}^{\mathrm{c}} = \{ (b); b \in \{0,1,\ldots,Y \} \}.
\end{equation}
 
The state space of secondary transmitter $n$ is denoted by 
\begin{equation}
	{\mathcal{S}}_n	=	\Big\{ (q_n, c_n) ; q_n \in \{0,1,\ldots,Q_n\}, c_n \in \{0,1,\ldots,C_n\} \Big\},
\end{equation}
where $q_n$ represents the queue state or data state, i.e., the number of data units in the data queue, and $c_n$ represents the energy state, i.e., the number of energy units in the energy storage. 

Now, we define the state of the blockchain environment. The mempool states of $K$ blockchain networks are broadcast in the network, and thus the state of the blockchain environment is composed of the mempool states of the blockchain networks. Denote $\mathcal{S}^k$ as the mempool state of blockchain network $k$. Assuming that every transaction submitted to the mempool is attached with a transaction fee rate $F$, where $F$ is within $[F_{\min}, F_{\max}]$. The transaction fee rate range is divided into $M$ equal intervals, i.e., $(\Delta F_1,\ldots,\Delta F_M)$. $\mathcal{S}^k$ refers to the current number of data units and the corresponding transaction fee rate intervals in the mempool of blockchain network $k$ and is defined by
\begin{equation}
	{\mathcal{S}}^k = \{(m^k_1, m^k_2, \ldots, m^k_M)\},
\end{equation}
where $m^k_i$ is the current number of data units of the transactions in blockchain network $k$ that have transaction fee rates within the interval of $\Delta F_i$.

Then, the state space of the network is defined by
\begin{equation}
	{\mathcal{S}}	=	{\mathcal{S}}^{\mathrm{c}} \times \prod_{n = 1}^N {\mathcal{S}}_n \times \prod_{k = 1}^K {\mathcal{S}}^k 	,
\end{equation}
where $\times$ and $\prod$ represent the Cartesian product.

\subsection{Action space}
The gateway needs to decide the time scheduling for the secondary transmitters, the blockchain network, and the transaction fee rate attached to the transaction. In particular for the time scheduling, the gateway decides the number of time slots for the energy harvesting, the data backscatter, and the active transmission. Thus, the action space is defined as follows:
\begin{eqnarray}
	{\mathcal{A}}	&  = &  	\Bigg\{ (\mu, \alpha_1,\ldots,\alpha_N, \beta_1,\ldots,\beta_N, \kappa , f );  \\ \nonumber 	
			& &	\mu + \sum_{n=1}^N  \alpha_n \leq b, \mu + \sum_{n=1}^N ( \alpha_n + \beta_n ) \leq Y,  \\ \nonumber 
	        & &  \kappa  \in \{1, 2, \ldots, K\}, f \in \{1, 2, \ldots, M\}\Bigg\},
\label{eq:actionspace}
\end{eqnarray}
where $\mu$ is the number of busy time slots for the energy harvesting for the secondary transmitters, $\alpha_n$ is the number of busy time slots assigned to secondary transmitter $n$ for the data transmission in the backscatter mode, and $\beta_n$ is the number of idle time slots assigned to secondary transmitter $n$ for the data transmission in the active mode. The constraint $\mu + \sum_{n=1}^N  \alpha_n \leq b$ ensures that the number of time slots for the energy harvesting and the data transmission in the backscatter mode do not exceed the number of busy time slots. The constraint $\mu + \sum_{n=1}^N ( \alpha_n + \beta_n ) \leq Y$ ensures that the number of time slots for the energy harvesting and the data transmissions in the backscatter and active modes do not exceed the total number of time slots of the frame. In (\ref{eq:actionspace}), $\kappa$ is the index that refers to blockchain network $\kappa$ selected by the gateway, and $f$ is the index that refers to the transaction fee rate interval $\Delta F_f$. For example, for $f=2$, this means that the gateway decides the transaction fee rate interval $\Delta F_2$. Based on the index, the transaction fee rate that is attached to the transaction is calculated by
\begin{equation}
F = F_{\min} + (f-1 + \eta  )\frac{F_{\max} - F_{\min}}{M},
\end{equation}
where $\eta \in [0,1]$  is pre-defined by the gateway to indicate the representative value of the transaction fee rate interval. For example, when $\eta=1/2$ and $f=2$, the transaction fee rate attached to the transaction is the middle value of interval $\Delta F_2$.

\subsection{State transition}

Now, we consider state transition of the network. In the busy channel period, the number of time slots assigned to secondary transmitter $n$ for the energy harvesting is $b(t) - \alpha_n$. Thus, after the busy channel period, the number of energy units in the storage of the secondary transmitter changes from $c_n$ to $c_n^{(1)}$ as follows:
\begin{equation}
	c_n^{(1)}=\min \big(c_n + (b(t) - \alpha_n)e^{\mathrm{h}}_n,C_n \big).
\end{equation}

Likewise, the number of data units in the data queue of secondary transmitter $n$ changes from $q_n$ to $q_n^{(1)}$ as follows:
\begin{equation}
	q_n^{(1)}	=	\max \big(0, q_n - \alpha_n d^{\mathrm{b}}_n \big).
\end{equation}

During the idle channel period, secondary transmitter $n$ requires $q_n^{(1)}/d^{\mathrm{a}}_n$ time slots to transmit $q_n^{(1)}$ data units. However, the secondary transmitter is only assigned with $\beta_n$ time slots for the data transmission. Thus, it actually transmits its data units in $\min (\beta_n, q_n^{(1)}/d^{\mathrm{a}}_n)$ time slots. 

At the end of the idle channel period, the energy state of secondary transmitter $n$ changes from $c_n^{(1)}$ to  $c'_n$ as follows:
\begin{equation}
	c'_n	=	\max \big(0, c_n^{(1)} - \min (\beta_n, q_n^{(1)}/d^{\mathrm{a}}_n)e^{\mathrm{a}}_n \big).
\end{equation}

Likewise, the number of data units in the data queue of secondary transmitter $n$ changes from $q_n^{(1)}$ to $q_n^{(2)}$ as follows:
\begin{equation}
	q_n^{(2)}	=	\max \big(0, q_n^{(1)} - \min (\beta_n, c_n^{(1)}/e^{\mathrm{a}}_n)d^{\mathrm{a}}_n \big).
\end{equation}

Note that at the end of each time frame, the secondary transmitters aggregate their sensing data and update the data to their queues. Assuming that the number of data units arriving at secondary transmitter $n$ during the time frame is $z_n$, which typically follows the Poisson distribution ${Pois(\lambda_n)}$~\cite{consul1973}. $\lambda_n$ is the data arrival rate, which indicates the average number of data units designated in a time frame. Then, the probability of $w$ data units arriving in secondary transmitter $n$ during the time frame is
\begin{equation}
Pr(z_n = w )= e^{-\lambda} \frac{\lambda^w}{w!}.
\end{equation}

As such, at the end of the time frame, the number of data units in the data queue of secondary transmitter $n$ changes from $q_n^{(2)}$ to $q'_n$ as follows:
\begin{equation}
	q'_n	=	q_n^{(2)} + z_n.
\end{equation}

The total number of data units that $N$ secondary transmitters transmit to the gateway in the time frame is
\begin{equation}
	D =	\sum_{n=1}^N S^{\mathrm{b}}_n (q_n^{(1)} - q_n) + \sum_{n=1}^N S^{\mathrm{a}}_n (q_n^{(2)} - q_n^{(1)}).
	\label{eq:data}
\end{equation}

The first and the second terms of (\ref{eq:data}) refer to the total numbers of data units transmitted in the backscatter mode and the active mode, respectively. Then, the gateway (i) includes $D$ data units into a packet, (ii) generates a transaction, and (iii) transmits the transaction to the mempool.

The transition of the mempool state from time frame $t$ to $t+1$ depends on (i) transactions arriving in the mempool, and (ii) mining process in which the miner adds unconfirmed transactions into the new block at time slot $t$. Note that apart from the transaction transmitted by the gateway, there are transactions transmitted, e.g., by other gateways or users in the network. We assume that there are $Z$ other transactions arriving in the mempool in every time frame. Each transaction has a data size and a transaction fee rate that follow the uniform distributions $U[1, D_{\max}]$ and $U[F_{\min}, F_{\max}]$, respectively. In the mining process, each miner attempts to collect transactions as many as possible from the mempool to form a new block. The new block is assumed to have a size of $B$ data units, $B \leq B_{\max}$. Then, the new mempool state is determined based on the remaining transactions in the mempool. In particular, the new mempool state is composed of the number of data units and the corresponding transaction fee rates in the mempool.

\subsection{Reward function}
The objective of the problem is to maximize the number of data units stored in the blockchain while minimizing the storage cost. Therefore, the reward function $\mathcal{R}$ of the gateway should consist of two components, i.e., the positive utility $R_{\text{success}}$, and the transaction fee $C_T$. The positive utility $R_{\text{success}}$ depends on the throughput $D_{th}$ that is defined as the number of data units from the gateway successfully stored in the blockchain in each time frame. The data units are considered to be successfully stored in the blockchain if the transaction including these data units is added to the new block at the current time frame and is not attacked. In particular, in time frame $t$, if the transaction containing $D$ data units is included in the new block and is not attacked, then $D_{th}=D$, and otherwise  $D_{th}=0$. Assuming that the gateway receives a reward of $\rho$ for each successfully stored data unit, then the positive utility that gateway receives during time frame $t$ is
\begin{equation}
    R_{\text{success}} = D_{th} \rho.
\end{equation}

The transaction fee $C_T$ depends on the number data units included in the transaction and the transaction fee rate $F$ that the gateway is willing to pay. $C_T$ is calculated as follows:
\begin{equation}
   C_T = D F.
\end{equation}

The objective is to maximize the utility of the gateway and the throughput of network while minimizing the transaction fee. Thus, the reward function of the network is defined as
\begin{equation}
   r = R_{\text{success}} - C_T.
   \label{eq:reward}
\end{equation}

In summary, the gateway observes the network to define the initial state $s_0 \in \mathcal{S}$. Given the current state $s \in \mathcal{S}$, the gateway executes an action $a \in \mathcal{A}$ that includes the time schedule and the transaction fee rate. The gateway then broadcasts the time schedule to the secondary transmitters in the network. Each secondary transmitter can harvest energy, backscatter its data or actively transmit its data according to the time schedule. The gateway generates the transaction and sends it to the mining pool. At the end of the time frame, the gateway receives the reward $r(s,a)$. Also, the network state transits to a new one $s' \in \mathcal{S}$. The problem of the gateway is to find the optimal policy $\pi ^*: \mathcal{S} \rightarrow \mathcal{A}$ to maximize the long-term reward $\Phi$ that is defined by
\begin{equation}
    \Phi = \sum_{t=0}^{T} \gamma^t {r}_t,
\end{equation}
where ${r}_t = \mathbb{E} \left [ r(s_t, a_t) \right ]$, $T$ is the length of the time horizon, and $\gamma$ is the discount factor for $0 \leq \gamma < 1$. The discount factor $\gamma$ determines the importance of future rewards. A factor of 0 will make the agent short-sighted by only considering current rewards. $\gamma$ should be set less than 1 to indicate that future rewards are worth less than immediate rewards.

To solve the stochastic problem of the gateway, the standard learning algorithms such as Q-learning~\cite{watkins1992q} are typically adopted. However, the Q-learning can efficiently solve the stochastic optimization problem as the state and action spaces are small. For our model, the problem is complicated with the large state and action spaces. Specifically, the gateway needs to update a multi-dimensional state which includes the channel state, the states of the secondary transmitters and the mempool state. Furthermore, the gateway needs to determine the joint time schedule, blockchain network and transaction fee rate decisions. Thus, the Q-learning may not be able to find the optimal policy. To overcome the shortcoming of the Q-learning, we propose to employ, evaluate, and assess the DRL with D3QN~\cite{wang2016} to find the optimal policy of the gateway.

\section{Deep Reinforcement Learning Algorithm}

Similar to the Q-learning, the DRL allows the gateway to map its state to an optimal action. However, a Deep Neural Network (DNN) with weights $\boldsymbol{\theta}$ is used to derive the approximate values $\mathcal{Q}^*(s,a)$ instead of the Q-table. The input of the DNN includes the state of the network, and the output includes Q-values $\mathcal{Q}(s,a;\boldsymbol{\theta})$ of all possible actions. $\mathcal{Q}(s,a)$ can be expressed by two elements as follows:
\begin{equation}
    \mathcal{Q}(s,a) = \mathcal{V}(s) + \mathcal{A}(s, a),
    \label{eq:Q_V_A}
\end{equation}
where $\mathcal{V}(s)$ is the value of state $s$, and $\mathcal{A} (s, a)$ is the advantage of taking action $a$ at state $s$. In particular,  $\mathcal{A} (s, a)$ shows how much better to take action $a$ compared with other possible actions at that state. As such, the DRL using the D3QN separates the estimator of the two elements by dividing the DNN into two streams: one that uses value stream parameters $(\boldsymbol{\theta}, \nu)$ to estimate the state value $\mathcal{V}(s)$, and one that uses advantage stream parameters $(\boldsymbol{\theta}, \xi)$ to estimate the advantage for each action $\mathcal{A} (s, a)$. By decoupling the estimation, $\mathcal{V}(s)$ is determined, and thus the DRL can learn valuable states without learning the effect of each action at each state. Then, these two streams are combined through a special aggregation layer to estimate $\mathcal{Q}(s,a)$ as follows:
\begin{align}
\label{eq:Q_D3QN_update}
	\mathcal{Q}(s,a;\theta) =  &  \mathcal{V}(s;\theta, \nu) 	\\
				& + \left ( \mathcal{A}(s,a; \theta, \xi) - \frac{1}{\left | \mathcal{A} \right |} \sum_{a'} \mathcal{A}(s,a'; \theta, \xi) \right ),\notag
\end{align}
where $\left | \mathcal{A} \right |$ is the dimension of vector $\mathcal{A}(s,a; \theta, \xi)$. Note that the advantage function estimator is forced to be zero at the chosen action by subtracting the average advantage of all possible actions of the state as in (\ref{eq:Q_D3QN_update}). This is to address the identifiability issue that is not able to find $\mathcal{V}(s)$ and $\mathcal{A} (s, a)$ given $\mathcal{Q}(s,a)$ by using (\ref{eq:Q_V_A}).

In general, the gateway executes action $a$ and receives experience $<s,a,r,s'>$. Based on the experiences, the gateway updates the weights of the DNN. In fact, when the Q-network is updated, the Q-values may change. This causes a high fluctuation of the value estimations. To stabilize the network performance, the DRL uses the D3QN that is composed of a target Q-network with weights $\boldsymbol{\theta}'$ and an online Q-network with weights $\boldsymbol{\theta}$. Weights $\boldsymbol{\theta}'$ of the target Q-network are updated by copying weights $\boldsymbol{\theta}$ of the online Q-network after every $L^-$ steps. The online Q-network updates its weights $\boldsymbol{\theta}$ at each iteration to minimize the loss function defined as 
\begin{equation}
L= \mathbb{E}\left[ \Big (y-\mathcal{Q}(s,a;\boldsymbol{\theta}) \Big)^2\right], 
\label{D3QL_loss}
\end{equation}
where $y$ is the target value. $y$ is determined by
\begin{equation}
y = r + \gamma \mathcal{Q}\Big{(} s', \arg\max_{a' \in \mathcal{A}} \mathcal{Q}(s',a';\boldsymbol{\theta});\boldsymbol{\theta}'\Big{)}.
\label{y_value_D3QL}
\end{equation}

Note that the expected value $y$ is calculated based on the immediate reward $r$, the discount factor $\gamma$ and the estimate of the optimal future value. The optimal future value is obtained from the weights $\boldsymbol{\theta}'$of the target network, and the selection of an action, in the $\arg\max$, is based on the online weights $\boldsymbol{\theta}$. As such, the DRL is able to prevent the overoptimistic estimation problem in which the same value is used to both decide the best action, i.e., with the highest expected reward, and estimate the action value.


To overcome the instability of the learning process, the experience replay strategy is used. In particular, a replay memory $\mathcal{D}$ is used to store experiences $<s,a,r,s'>$. The online Q-network is trained by using mini-batches of experiences that are randomly sampled from the replay memory.  The Q-values obtained by the online Q-network are used to select actions, e.g., through the $\epsilon$-greedy policy, and obtain the new experiences. The experiences are then stored in the replay memory. The experiences in the replay memory are independent and identically distributed, and thus the correlation among the training examples is reduced. Therefore, the experience replay strategy ensures that the optimal policy cannot be driven to a local minima.

\begin{algorithm}
\small
\caption{The D3QN algorithm for time scheduling, blockchain selection and transaction fee rate decisions of the gateway.} \label{D3QL_Algorithm}

\hspace*{\algorithmicindent} \textbf{Input:} Action space $\mathcal{A}$, number of training episodes $T_e$, number of steps in one episode $T_s$, discounted factor $\gamma$, mini-batch size $B$, target network replacement frequency $F_r$;

\hspace*{\algorithmicindent} \textbf{Output:} Optimal policy $\pi^*$.

\begin{algorithmic}[1]

\State \textbf{Initialize:} Replay memory $\mathcal{D}$, online network with random weights $\boldsymbol{\theta}$, target network with weights $\boldsymbol{\theta'} = \boldsymbol{\theta}$.

\For {\text{episode $i = 1$ to $T_e$}}:

\State Initialize network state $s$.

\For {\text{step $j = 1$ to $T_s$}}

\State Choose action $a$ according to $\epsilon-greedy$ from $\mathcal{Q}(s,a; \boldsymbol{\theta})$.

\State Broadcast scheduling massages to secondary transmitters.

\State Receive data, compress into packet, submit transaction to mining pool of selected blockchain network, and calculate immediate reward $r$.

\State Receive state massages from primary transmitter and $N$ secondary transmitters, observe mempool of all connected blockchain networks to update new state $s'$.

\State Store tuple $<s, a, r, {s}'>$ in $\mathcal{D}$.

\State Sample $H$ experiences $e_k = <s_k, a_k, r_k, {s}'_k>$ from $\mathcal{D}$.

\For{\text{$k=1$ to $H$}}

 \State Calculate 
 \begin{equation}
y_k = r_k + \gamma \mathcal{Q}\Big{(} s'_k, \arg\max_{a' \in \mathcal{A}} \mathcal{Q}(s'_k,a';\boldsymbol{\theta});\boldsymbol{\theta}'\Big{)}. \notag
\end{equation}

\EndFor 
 
\State Define $\overline{L}=\frac{1}{H} \sum_{k=1}^{H} \Big{(}y_k -\mathcal{Q}(s_k,a_k;\boldsymbol{\theta}) \Big{)} ^{2}$.

\State Update $\boldsymbol{\theta}$ by performing a gradient descent step on $\overline{L}$.

\State Reset $\boldsymbol{\theta'} = \boldsymbol{\theta}$ every $F_r$ steps.

\State Set $s\leftarrow {s}'$.

\EndFor
\EndFor

\end{algorithmic}
\end{algorithm}

Algorithm \ref{D3QL_Algorithm} shows how to implement the D3QN algorithm to achieve the optimal time schedule, blockchain network selection, and transaction fee rate decision in the RF-power backscatter  cognitive  radio  based  blockchain  network. In the algorithm, each step $j$ corresponds to a time frame. The algorithm is implemented in two phases, i.e., the experience phase and the training phase. In the experience phase, the gateway performs an action $a \in \mathcal{A}$ according to the $\epsilon$-greedy policy. At the end of the time frame, the gateway records experience $e=<s,a,r,s'>$ the replay memory. In the training phase, the algorithm takes a mini-batch of $H$ experiences $e_k = <s_k, a_k, r_k, s'_k>$ from $\mathcal{D}$ to train the online Q-network. The target value $y_k$ corresponding to experience $e_k$ is calculated by using both the online Q-network and target Q-network according to (\ref{y_value_D3QL}). Then, the mean loss $\overline{L}$ over the mini-batch is determined according to (\ref{D3QL_loss}). Weights $\boldsymbol{\theta}$ of the online network are adjusted by performing the gradient descent step on $\overline{L}$. Weights $\boldsymbol{\theta'}$ of the target network are updated by setting $\boldsymbol{\theta}=\boldsymbol{\theta'}$ in every $F_r$ steps.

\section{Performance Evaluation}
In this section, we present simulation results to evaluate the performance of the proposed DRL algorithm.
\subsection{Parameter setting}

Simulation parameters for the blockchain-based RF-powered backscatter cognitive radio network are shown in Table \ref{table:evironment_parameters}. In particular, each time frame consists of $7$ time slots. The gateway performs the allocation of time slots to the secondary transmitters. The maximum data queue size of each secondary transmitter is set to $7$ units, and the energy storage capacity is set to $5$ units. The number of blockchain networks is $K=3$.

\begin{table}[!h]
\caption{Parameters of the backscatter cognitive radio network and the blockchain networks}
\centering
\begin{tabular}{lc}
\hline\hline

{\em Parameters} 			& {\em Values} \\ [0.5ex]
\hline
Number of time slots in a time frame ($F$)    & 7 \\ 
Number of busy time slots in a time frame ($b(t)$) & ${[}1, 6{]} $   \\ 
Data queue size       ($Q_n$)    & 7   \\ 
Data state				($q_n$)		& $[0, Q_n]$ \\
Energy storage capacity ($C_n$)    & 5   \\ 
Energy state			($c_n$)		& $[0, C_n]$\\
Data arrival rate     ($\lambda_n $)  & ${[}1, 10{]} $ \\
$(e_h, d_b, e_t, d_t)$  & $(1, 1, 1, 2)$ \\
Mempool size  ($M_{\max}$)    & 50   \\ 
Maximum size of a block	($B_{\max}$)	& 30		\\
Number of transaction fee intervals ($M$)    & 4   			\\ 
${[} F_{\min}, F_{\max}{]}$					&  ${[}0.01, 0.8{]}$\\
$(\rho,\eta)$            & $(1,0.2)$   \\
$q$            & $[0, 0.1]$   \\
\hline
\end{tabular}
\label{table:evironment_parameters}
\end{table}

\begin{table}[!h]
\caption{Parameters of the algorithms}
\centering
\begin{tabular}{ll}
\hline\hline
{\em Parameters} 			& {\em Values} \\ [0.5ex]
\hline
Fully connected neuron network size & 32x32x32            \\
Activation                          & ReLU                \\
Optimization algorithm     			& Adam                \\
Learning rate ($\delta$)            & 0.001               \\ 
Discount rate ($\gamma$)            & 0.9                 \\ 
$\epsilon$                 & 0.9 $\rightarrow$ 0 \\ 
Mini-batch size ($H$)                    & 32                  \\
Replay memory size			& 50000               \\ 
Number of steps in one episode ($T_s$)         & 200                 \\
Number of training episodes ($T_e$)             & 50000             \\
Target network replacement frequency ($F_r$) & 10000               \\
\hline
\end{tabular}
\label{table:system_parameters}
\end{table}

Table \ref{table:system_parameters} lists parameters for implementing the DRL algorithm. These parameters are similar to those in other works~\cite{anh2019}, \cite{nguyen2019}. Accordingly, the DRL algorithm is implemented by using the TensorFlow deep learning library. The learning rate is set to low, i.e., 0.001, to ensure that the training phase does not miss local minima. However, this may slow down the training progress. To achieve the fast and smooth convergence, we use the
Adam optimizer that allows to adjust the learning rate during the training phase. The training data includes past experiences of the gateway that are stored in replay memory $\mathcal{D}$. The gateway can receive its experience $(s,a,r_k,s')$ by executing an action $a$. The $\epsilon-greedy$ policy is used for executing actions. Here, we set $\epsilon= 0.9$ that balances the exploration and exploitation. This means that a random action is selected with a probability of $\epsilon= 0.9$, and the best action, i.e., the action that maximizes the Q-value, is selected with a probability of $1-\epsilon= 0.1$. To move from a more explorative policy to a more exploitative one, the value of $\epsilon$ is linearly reduced from 0.9 to 0 during the training phase. The DRL algorithm is expected to obtain the long-term reward, and thus the discount factor is set to 0.9.

To evaluate the proposed DRL algorithm, we introduce four baseline schemes as follows:\\
$\circ$ \emph{Q-learning~\cite{watkins1992q}}: This scheme is also known as reinforcement learning algorithm.\\
$\circ$ \emph{Harvest-then-transmit (HTT)~\cite{lyu2018b}:} Each secondary transmitter harvests energy as the primary channel is busy, and the secondary transmitters transmits data as the channel is idle.\\
$\circ$ \emph{Backscatter policy~\cite{liu2013}:} Each secondary transmitter only performs backscattering as the primary channel is busy.\\
$\circ$ \emph{Random policy:} The gateway assigns randomly time slots to the secondary transmitters for the energy harvesting and data backscatter as the primary channel is busy. As the channel is idle, the time slots are randomly assigned to the secondary transmitters for the active data transmissions. \\

Note that in the HTT, Backscatter, and Random policies, the transaction fee rates  attached with the transactions may not be optimized. Instead, the gateway determines the transaction fee rates by using common mechanisms such as Bitcoin transaction fee estimation \cite{shehabi2018}. The main idea of the mechanism is that the transaction fee rate is calculated according to the average transaction fee rate of the last block of the main block chain.

\subsection{Numerical Results}

In this section, we provide the performance comparisons between the proposed DRL algorithm and the baseline schemes in different scenarios. In particular, the convergence comparison is shown in Fig.~\ref{convergence}, the performance comparisons when the number of busy time slots per time frame varies are shown in Figs.~\ref{busy_slot_throughput}, \ref{busy_slot_fee}, \ref{timeframe_throughput}, and \ref{timeframe_fee}, the performance comparison when the packet arrival probability is varied is shown in Fig. \ref{data_rate}, the performance comparisons when the probability that a block is found by the attacker are shown in Figs.~\ref{hastrate_throughput} and~\ref{hashrate_fee}, and the performance evaluations of the DRL algorithm when the number of blockchain networks varies are shown in Fig.~\ref{fig:multiblockchain}. Before presenting the performance comparisons among the algorithms, we discuss how the DRL algorithm obtains the optimal policy to maximize the long-term reward. 

\subsubsection{Optimal policy obtained by the DRL algorithm}
\begin{figure}[]
 \centering
\includegraphics[width=8.0cm, height = 6cm]{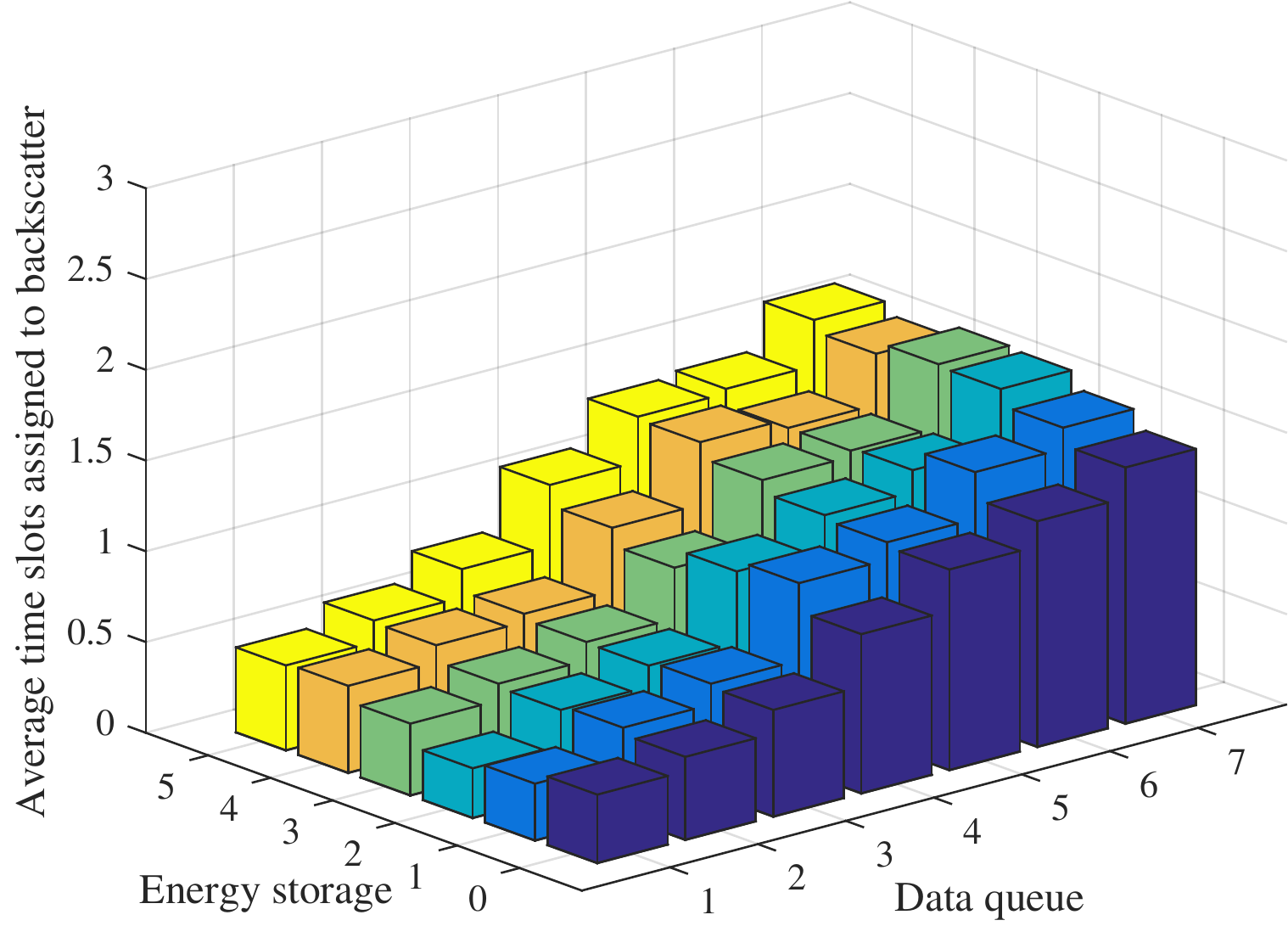}
 \caption{Time slots assigned to secondary transmitter 1 for the backscatter.}
  \label{backscatter}
\end{figure}

\begin{figure}[h]
 \centering
\includegraphics[width=8.0cm, height = 6cm]{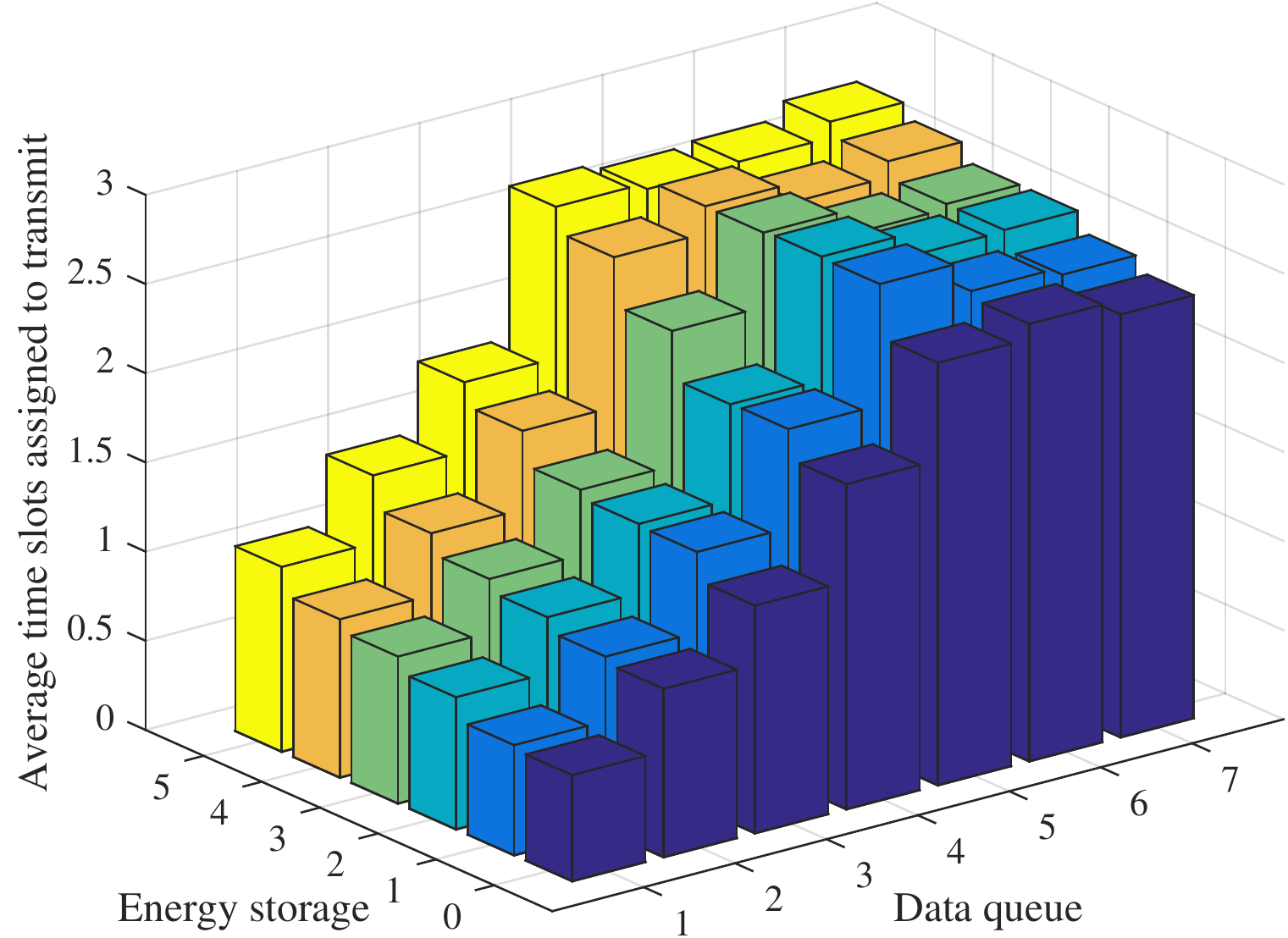}
 \caption{Time slots assigned to secondary transmitter
1 for the active transmission.}
  \label{transmit}
\end{figure}

With the DRL algorithm, the total reward depends heavily on the time scheduling policy of the gateway. This means that to achieve the high total reward, the gateway needs to take proper actions, e.g., assigning the number of time slots to the secondary transmitters for the data backscatter, data transmission, and energy harvesting. Thus, it is worth considering how the gateway takes the optimal actions for each secondary transmitter given its state. Without loss of generality, we consider the average number of time slots that the gateway assigns to secondary transmitter 1 for the data backscatter (Fig. \ref{backscatter}) and the data transmission (Fig. \ref{transmit}).

As shown in  Fig. \ref{backscatter}, the average number of time slots assigned to secondary transmitter 1 for the backscatter increases as its data queue increases. The reason is that as the data queue is large, the secondary transmitter needs more time slots to backcastter its data units. Thus, the gateway assigns more time slots to the secondary transmitter to maximize the throughput. It is also observed from Fig. \ref{backscatter} that the average number of time slots assigned to secondary transmitter 1 for the backscatter increases as its energy state increases. The reason is that as the energy state of the secondary transmitter is already high, the gateway assigns fewer time slots for the energy harvesting and prioritizes more time slots for the backscatter to improve the network throughput.

A high energy state allows the secondary transmitter to transmit more data units in the active transmission. Therefore, the gateway should assign more times slots to the secondary transmitter for the active data transmission. This is consistent with the simulation results shown in Fig. \ref{transmit}. Accordingly, the average number of time slots assigned to the secondary transmitter increases as its energy state increases. The figure also shows that as the data queue increases, the gateway assigns more data units to the secondary transmitter. This is evident since more data units in the data queue require more time slots for the active transmission.  

\subsubsection{Reward and convergence comparison}

\begin{figure}[]
 \centering
\includegraphics[width=7.5cm, height = 6cm]{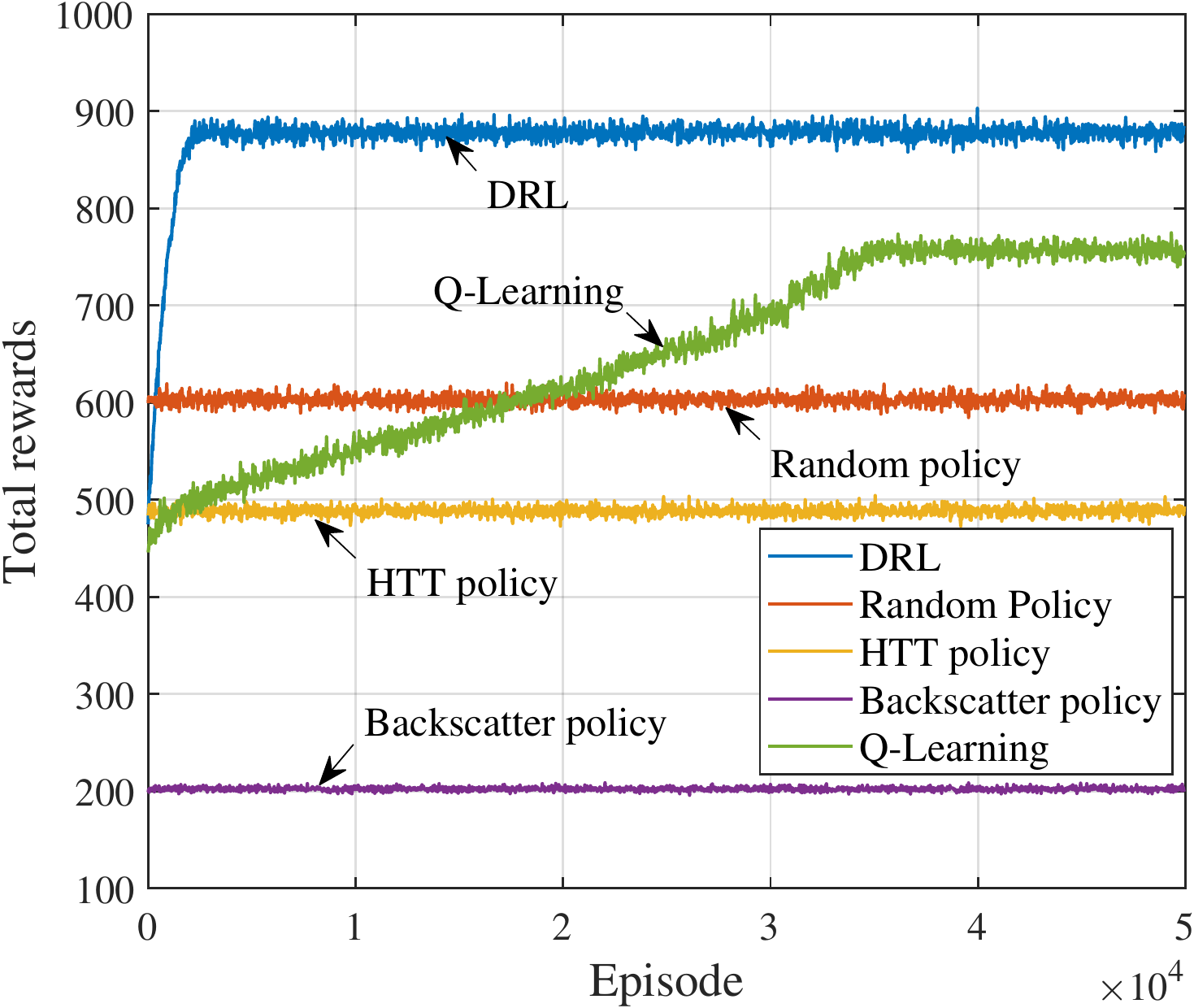}
 \caption{Convergence comparison.}
  \label{convergence}
\end{figure}

Next, we compare rewards obtained by the algorithms as well as the convergence speed of the algorithms. Fig.~\ref{convergence} illustrates the rewards obtained by the DRL, Random, HTT, Backscatter and Q-learning schemes. In fact, to enable the Q-learning scheme to run in our computation environment, we reduce the number of blockchain to $K=1$, and we reduce the state and action spaces by setting the following parameters: $F=5$, $Q_n=5$, and $E_n=3$. Moreover, we reduce the mempool size to $15$ and the size of the block to $10$. As seen, the DRL scheme converges to the reward that is much higher than those of the baseline schemes. In particular, the reward obtained by the DRL is around $880$, while those obtained by the Q-learning, Random, HTT, and Backscatter schemes are $757$, $603$, $488$ and $202$, respectively. As such, the reward obtained by the Q-learning scheme is lower than that obtained by the DRL scheme. The reason is that the state and action spaces are still too large for the Q-learning to update the whole Q-table. The Q-learning is thus not able to find the optimal policy. Moreover, the convergence speed of the DRL scheme is much faster than that of the Q-learning scheme. Specifically, the DRL scheme converges within $2500$ episodes, while the Q-learning scheme converges within $35000$ episodes. 

\subsubsection{Performance comparison}

We evaluate the proposed DRL scheme and compare the performance achieved by the proposed DRL scheme with those achieved by the baseline schemes by varying simulation parameters. Note that the parameters such as the data queue size, energy capacity, and the number of blockchain networks are set as in Table 1. Given the parameters, the Q-Learning algorithm cannot be applied in our computation environment due to the high complexity of the problem.

\textit{(a) Impact of the number of busy time slots per frame:}

\begin{figure}[]
 \centering
\includegraphics[width=7.5cm, height = 6cm]{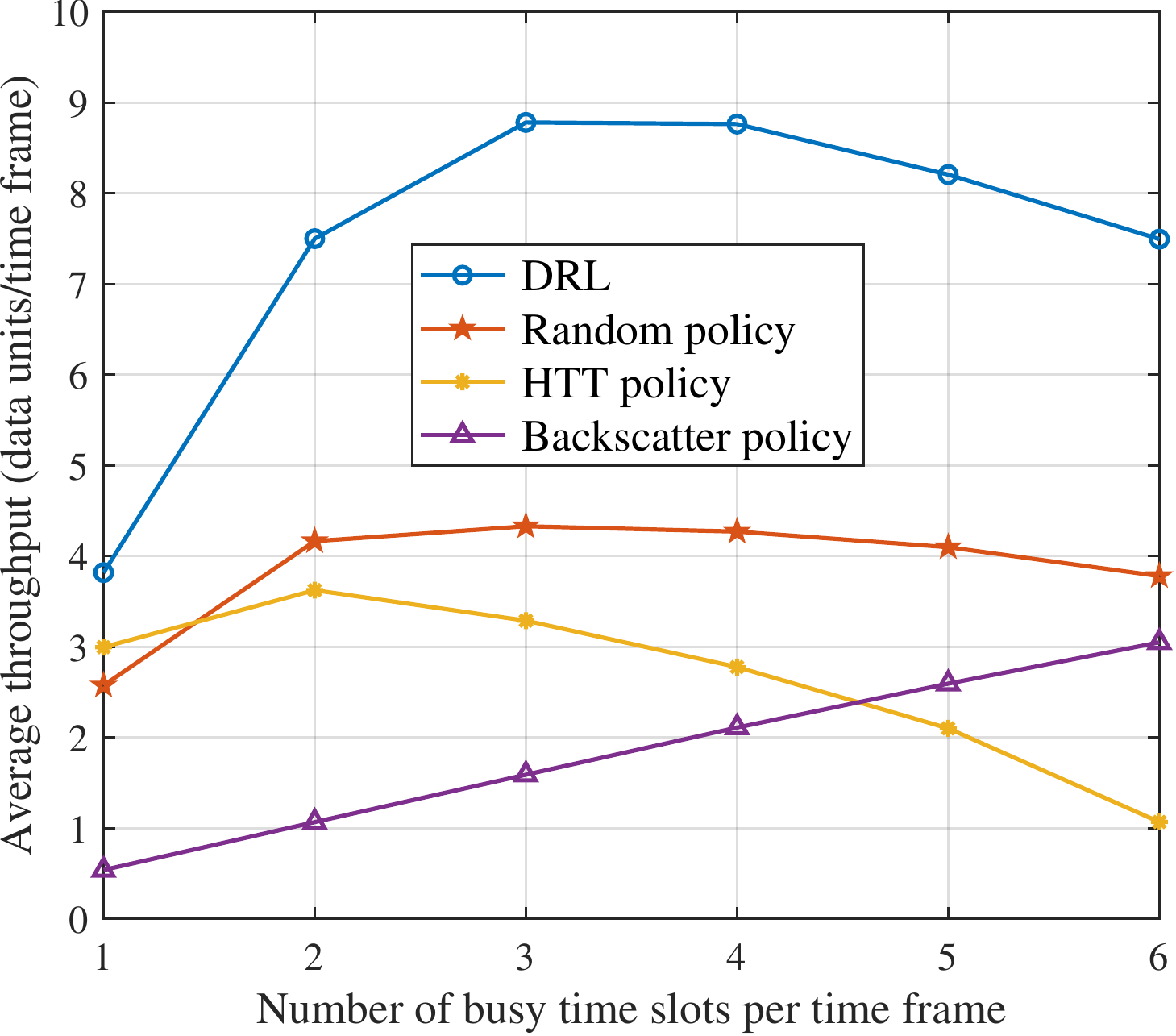}
 \caption{Average throughput versus the number of busy time slots.}
  \label{busy_slot_throughput}
\end{figure}

Figure~\ref{busy_slot_throughput} shows the performance comparison among the schemes as the number of busy time slots per frame is varied. As shown in the figure, the proposed DRL scheme significantly improves the throughput compared with the baseline schemes. Specifically, as the number of busy time slots per time frame is 3, the proposed DRL scheme improves the throughput up to $452.4\%$, $167.1\%$ and $102.9\%$ compared with the Backscatter, HTT, and Random schemes, respectively. In particular for the DRL, Random, and HTT schemes, the average throughput initially increases and then decreases as the number of busy time slots per time frame increases. The reason is that as the number of busy time slots per time frame is small, i.e., 1 and 2, the secondary transmitters favor the energy harvesting for the active data transmission during the idle channel period. However, if the busy channel period is long, i.e., the idle channel period is short, the secondary transmitters may have a low opportunity to transmit data from their data queues. 

\begin{figure}[]
 \centering
\includegraphics[width=7.5cm, height = 6cm]{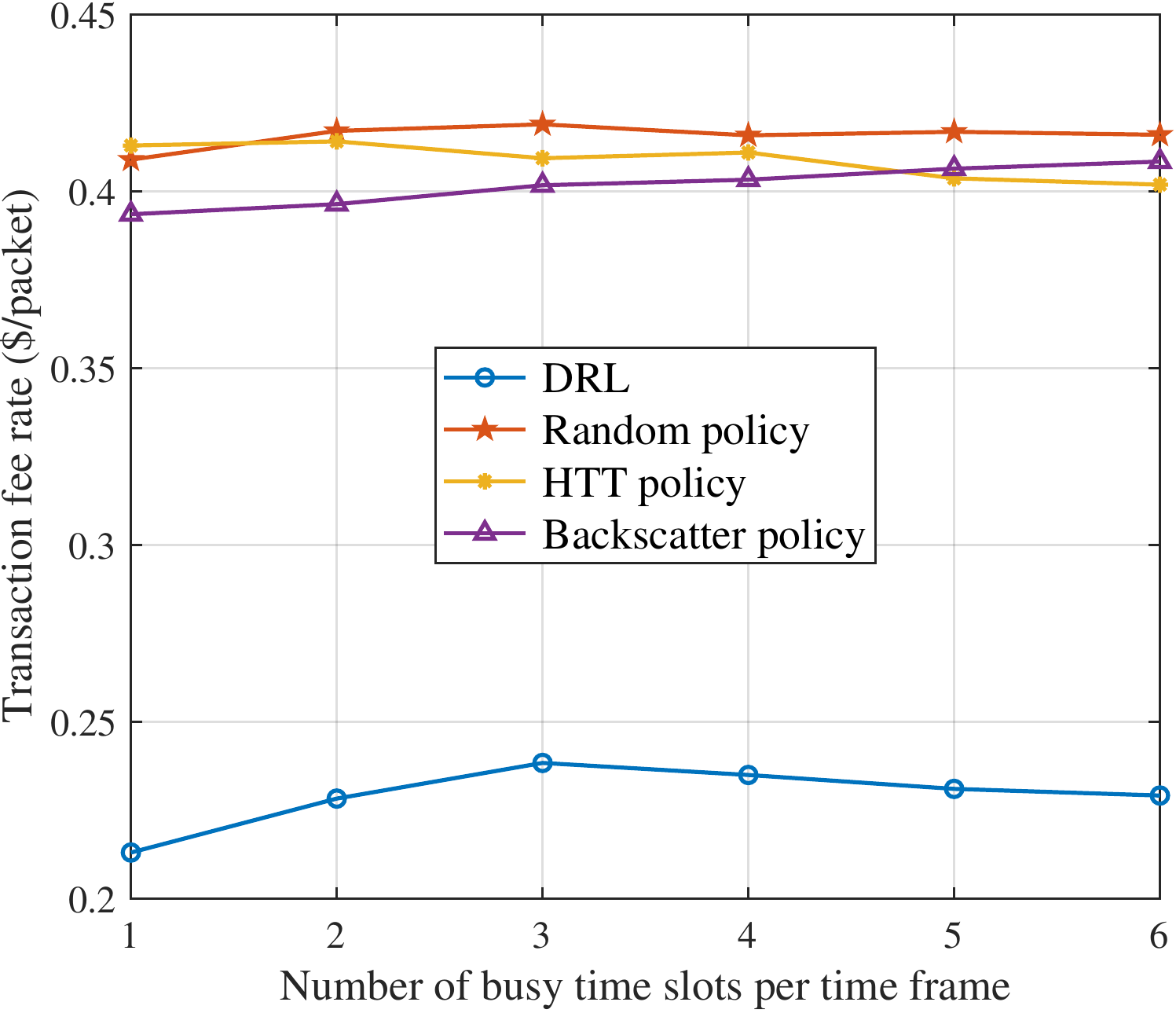}
 \caption{Transaction fee rate versus the number of busy time slots.}
  \label{busy_slot_fee}
\end{figure}

Next, we compare the transaction fee rate of the proposed DRL scheme and the baseline schemes as the number of busy time slots per time frame increases. Here, the transaction fee rate refers to the average cost that one data unit needs to be successfully stored in the blockchain. Thus, the transaction fee rate is expected to be low. As shown in Fig.~\ref{busy_slot_fee}, the transaction fee rate of the proposed DRL scheme is significantly improved compared with those of the baseline schemes. In particular, the transaction fee rate of the proposed DRL scheme is $75.3\%$ smaller than that of the Backscatter scheme, $78.4\%$ smaller than that of the HTT scheme, and $81.4\%$ smaller than that of the Random scheme. Note that the transaction fee rate of the DRL scheme slightly increases as the number of busy time slots per time frame increases, e.g., from 1 to 3. The reason can be explained as follows. As the number of busy time slots increases, the transaction that is sent from the gateway to the blockchain should include more data units. Given the constraint of the block size, the transaction has a lower probability to be added to the block. Thus, the gateway should choose a high transaction fee rate to improve the opportunity for the transaction to be successfully added to the block.

\textit{(b) Impact of the number of time slots per frame:}

\begin{figure}[]
 \centering
\includegraphics[width=7.5cm, height = 6cm]{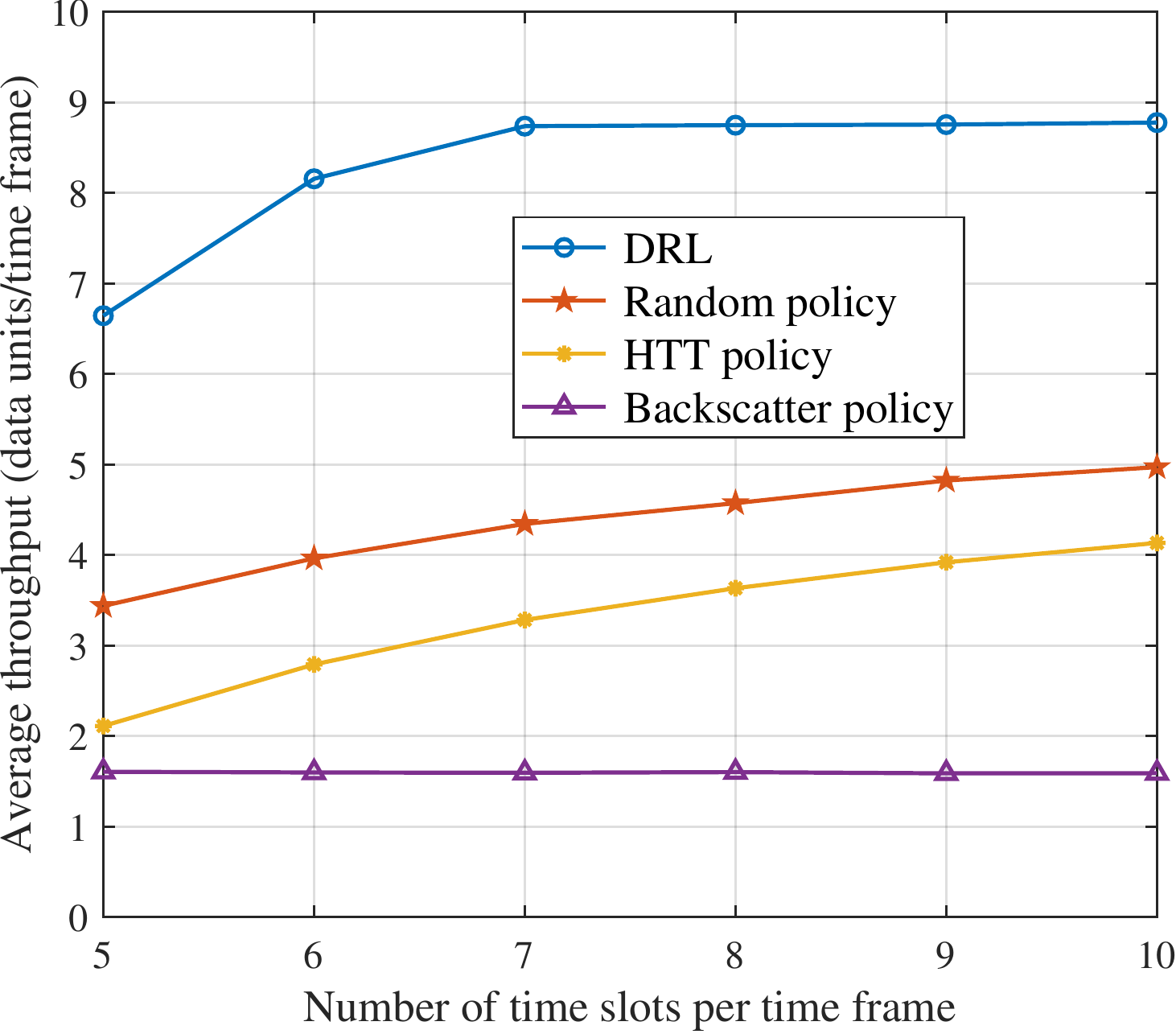}
 \caption{Average throughput versus the number of time slots per time frame.}
  \label{timeframe_throughput}
\end{figure}

As shown in Fig. \ref{timeframe_throughput}, the proposed DRL scheme always outperforms the baseline schemes in terms of throughput as the number of time slots per time frame varies. Moreover, the average throughput of the proposed DRL, Random, and HTT schemes increases with the increase of the frame length. The reason is that the secondary transmitters have more time slots for transmitting their data. In particular for the Backscatter scheme, the throughput remains stable with the increase of the frame length. This is because of that the data is only backscattered by secondary transmitters in busy channel period, and the busy channel period is fixed when increasing the frame length.

\begin{figure}[h]
 \centering
\includegraphics[width=7.5cm, height = 6cm]{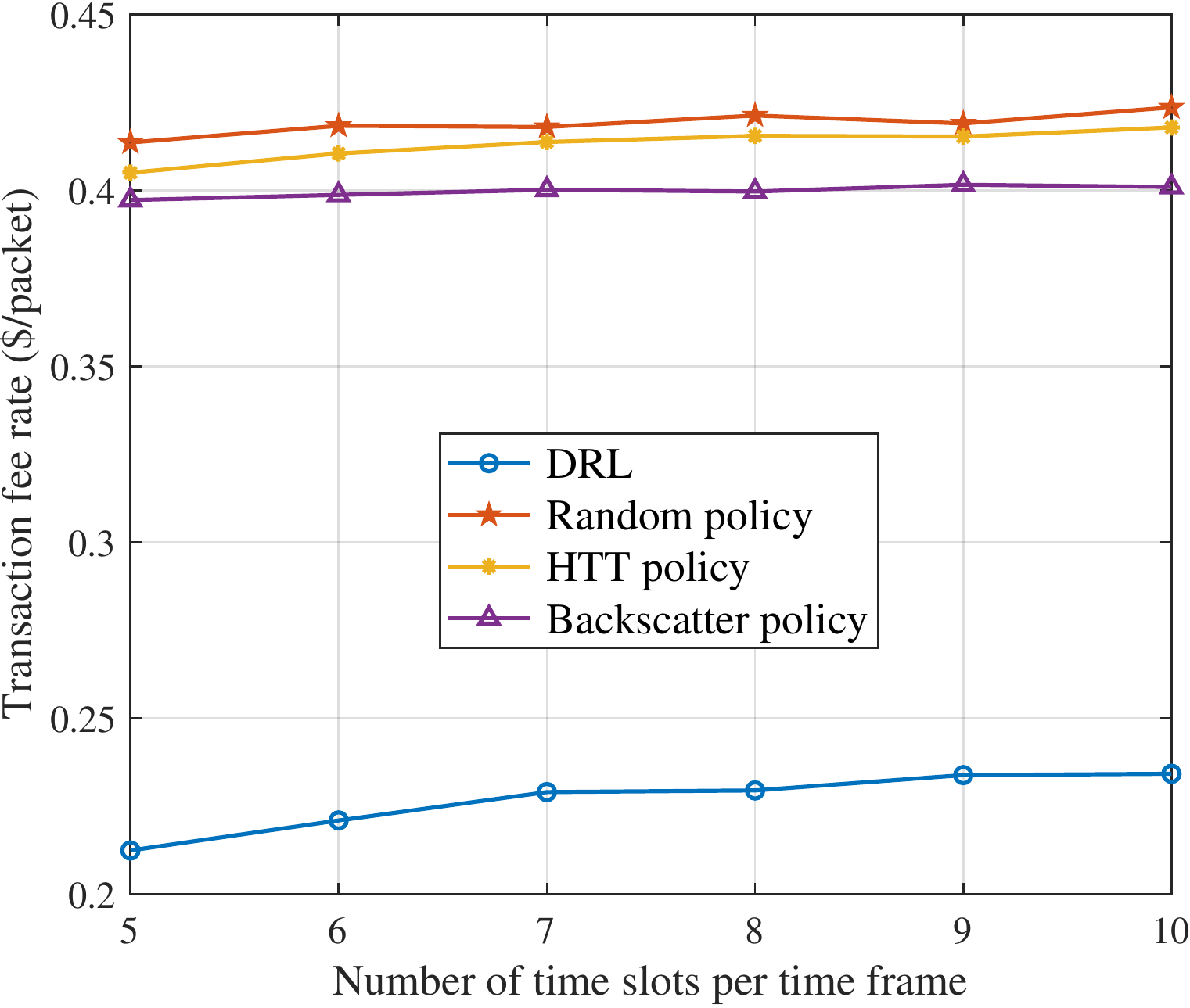}
 \caption{Transaction fee rate versus the number of time slots per time frame.}
  \label{timeframe_fee}
\end{figure}

Note that as the frame length increases, the transaction fee rates of the proposed scheme and the baseline schemes generally increase as shown in Fig.~\ref{timeframe_fee}. The reason is that as the frame length increases, the data packet sent to the blockchain may have a larger size. To have a higher opportunity to be added in the block, the gateway should choose a higher transaction fee rate. 

\textit{(c) Impact of the data arrival rate:}

\begin{figure}[h]
 \centering
\includegraphics[width=7.5cm, height = 6cm]{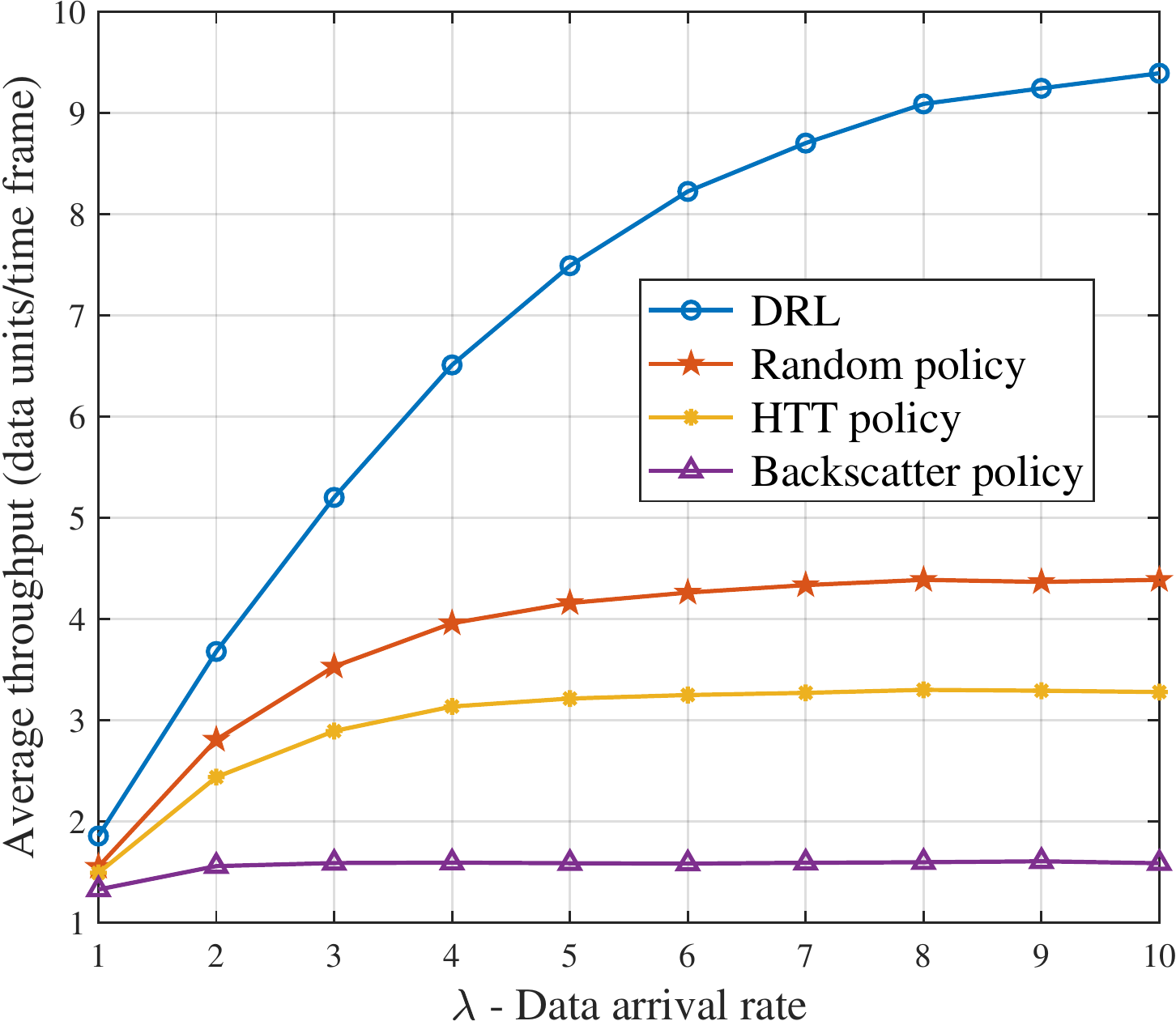}
 \caption{Average throughput versus the data arrival rate.}
  \label{data_rate}
\end{figure}

The performance improvement of the proposed DRL scheme compared with the baseline schemes is maintained when varying the data arrival rate. As shown in Fig. \ref{data_rate}, the average throughput obtained by the proposed DRL scheme is significantly higher than those obtained by the baseline schemes. For example, given a data arrival rate of $8$, the average throughput obtained by the proposed DRL scheme is up to $9.1$ data units per time frame, while those obtained by the Random, HTT, and Backscatter schemes are $4.4$, $3.3$, and $1.6$ data units per time frame, respectively. The gap between the proposed DRL scheme and the baseline schemes becomes larger as the data arrival rate increases.


\textit{(d) Impact of the blockchain network environments:}


\begin{figure}[]
 \centering
\includegraphics[width=7cm, height = 6cm]{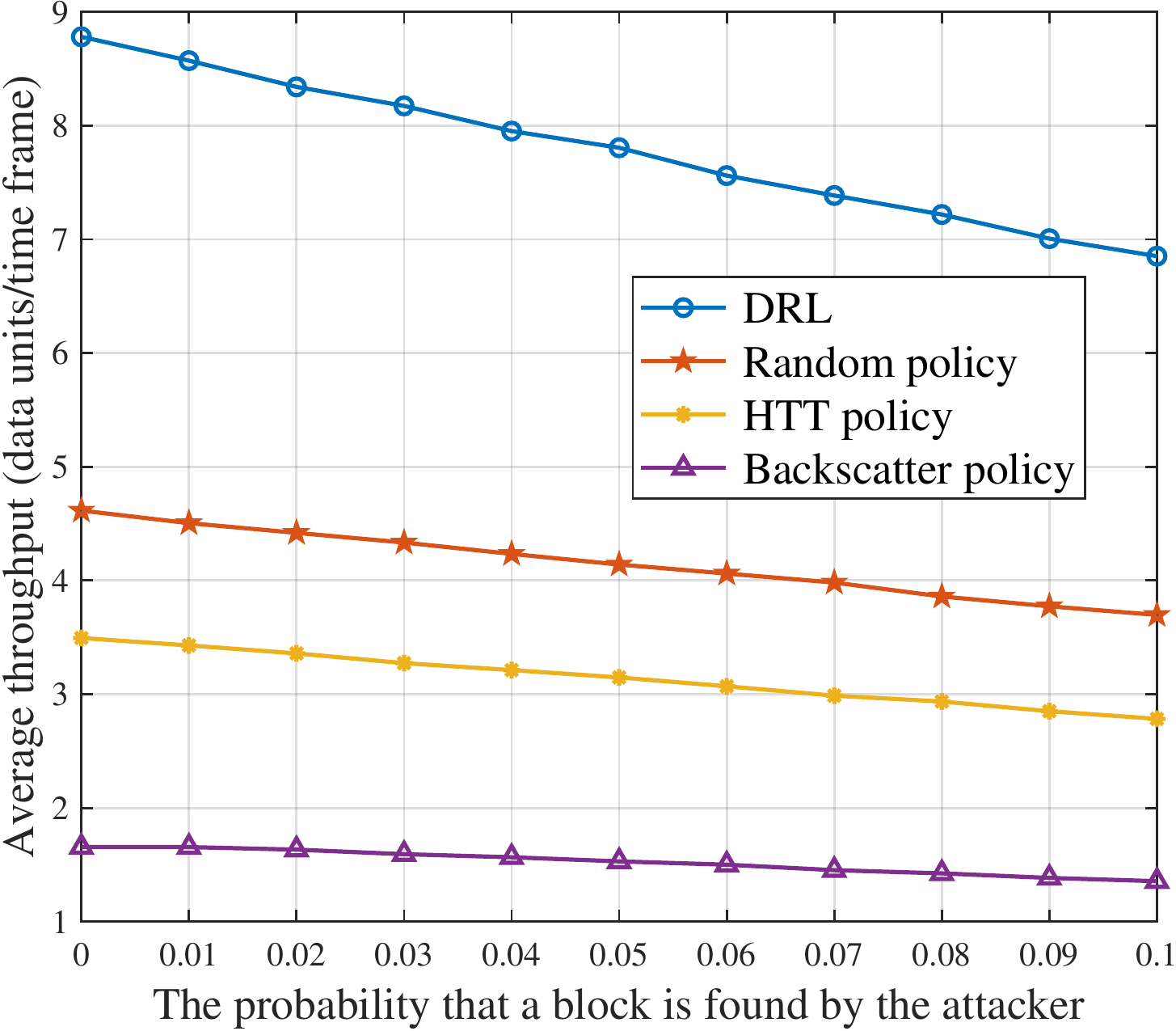}
 \caption{Average throughput versus the probability that a new block found by attacker.}
  \label{hastrate_throughput}
\end{figure}

\begin{figure}[h]
 \centering
\includegraphics[width=7cm, height = 6cm]{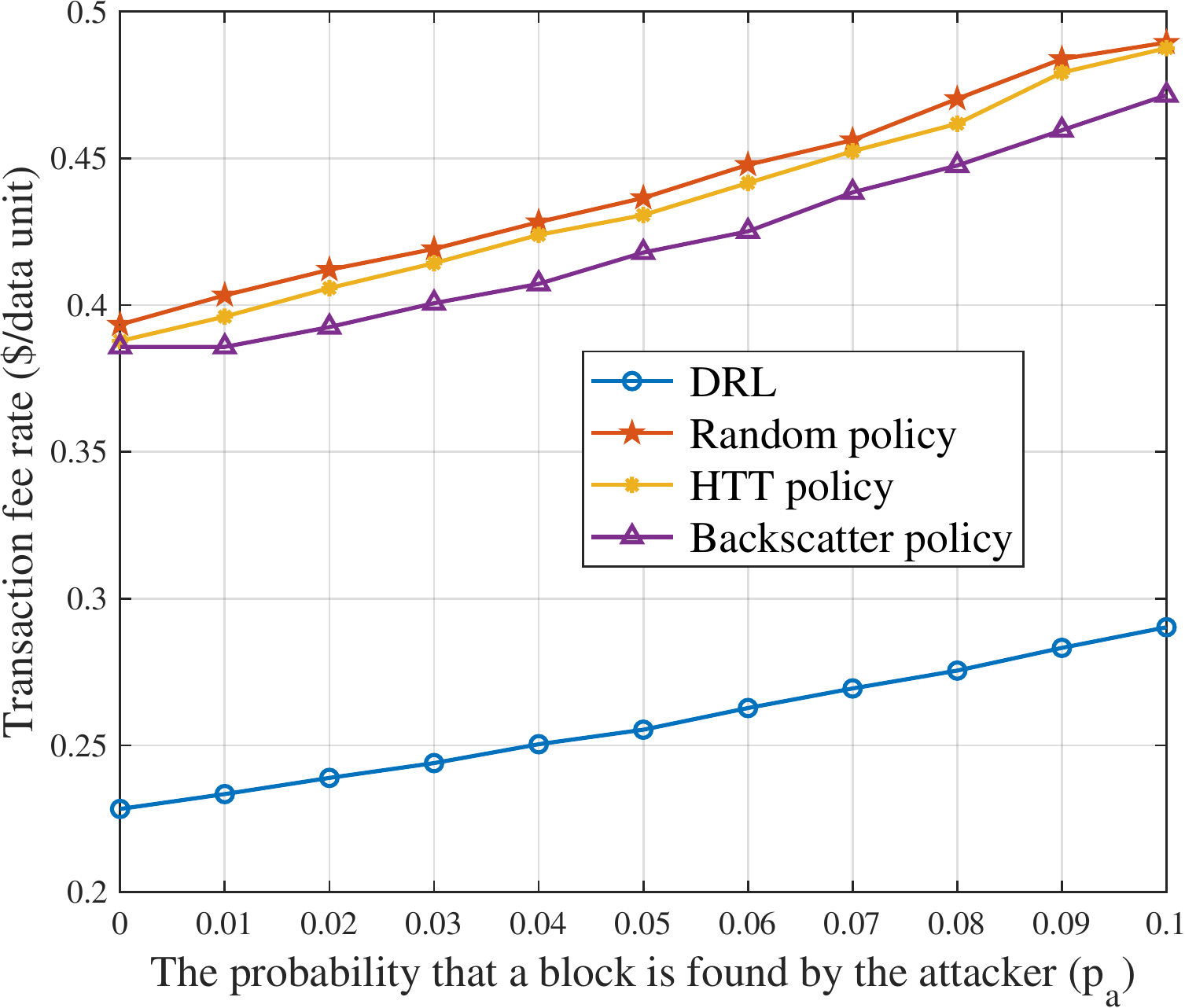}
 \caption{Transaction fee rate versus the probability that new block found by attacker.}
  \label{hashrate_fee}
\end{figure}

Next, we compare the performance obtained by the DRL scheme and that obtained by the baseline schemes as the probability $q$ that a block is found by the attacker varies. For the comparison purpose, we assume that the probability $q$ in the different blockchain networks is the same. As shown in Fig.~\ref{hastrate_throughput}, compared with the baseline schemes, the proposed DRL scheme still achieves the highest average throughput. It is also seen from the figure that as $q$ increases, the average throughputs of the schemes decrease significantly. The reason is that more transactions are attacked. This may incur a high cost due to storing transactions in the attacked blocks. As a result, the transaction fee rate rises significantly as shown in Fig. \ref{hashrate_fee}. 

\textit{(d) Impact of the number of blockchain networks:}
\begin{figure*}[!t]
\centering
\minipage{0.333\linewidth}
  \includegraphics[width= \textwidth]{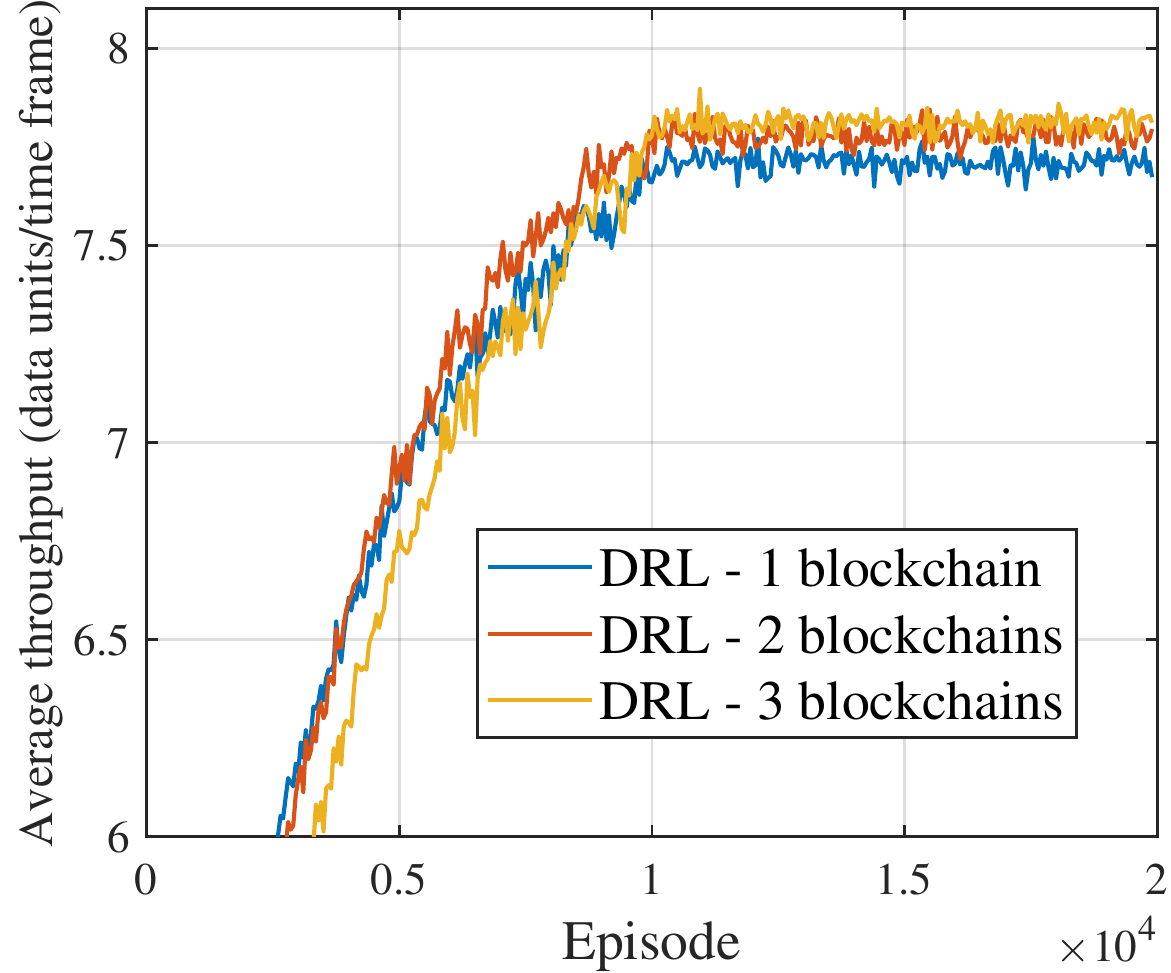}
  \subcaption{}
\endminipage
\minipage{0.333\linewidth}
  \includegraphics[width= \textwidth]{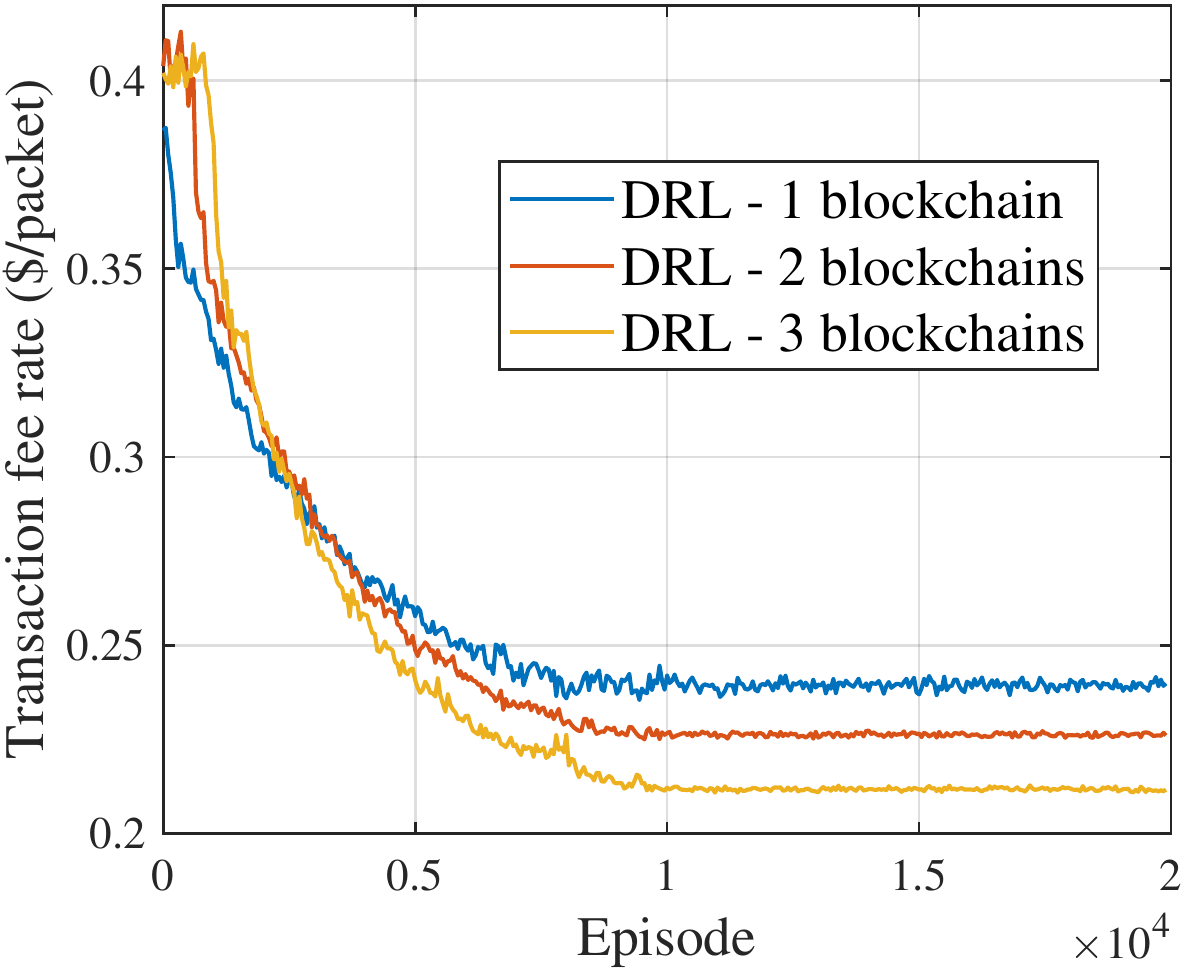} 
  \subcaption{}
\endminipage
\minipage{0.333\linewidth}
  \includegraphics[width= \textwidth]{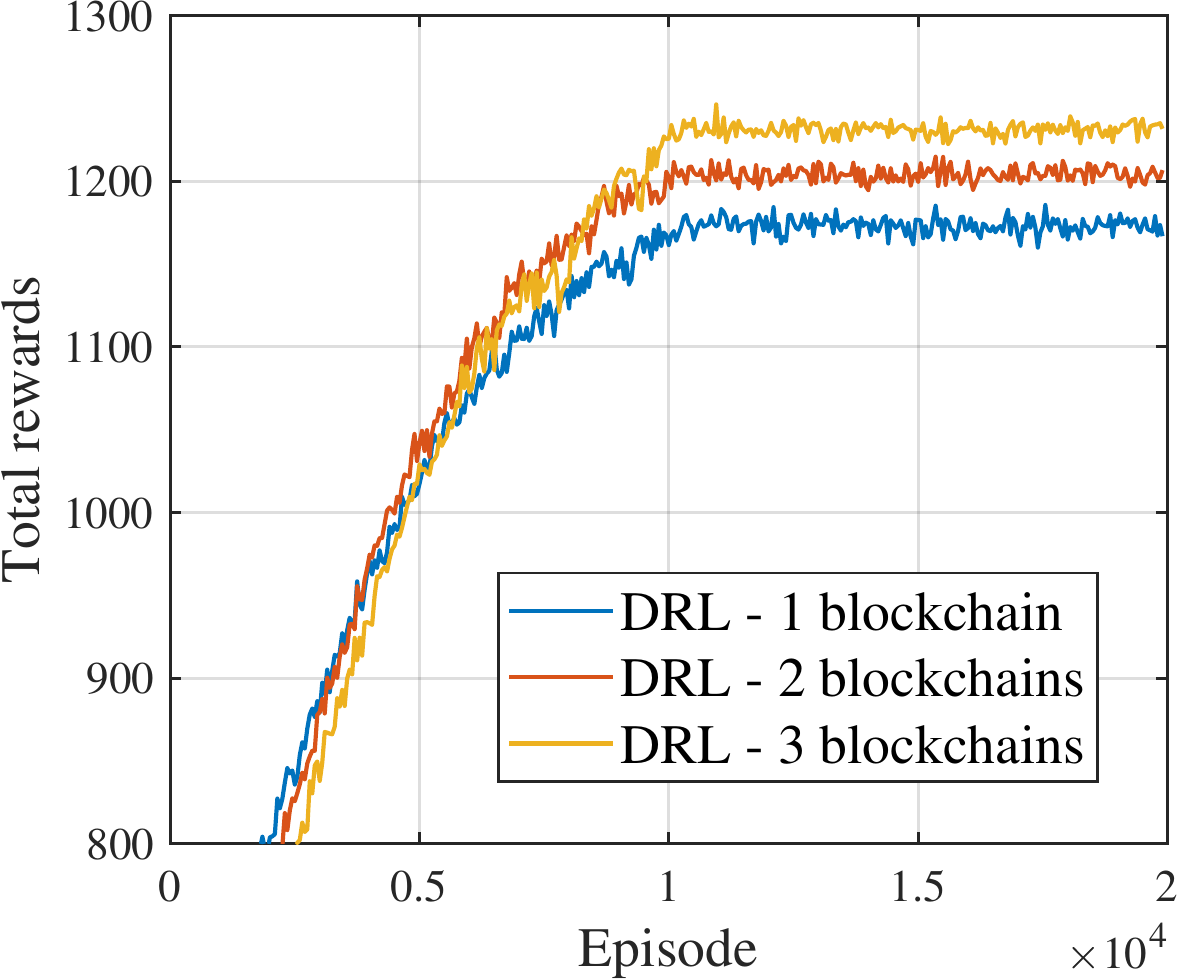}
  \subcaption{}
\endminipage
\caption{DRL performance in terms of (a) average throughput, (b) transaction fee rate, and (c) total reward.}
\label{fig:multiblockchain}
\end{figure*}
Finally, we evaluate the performance of the proposed DRL scheme when the number of blockchain networks $K$ varies. As seen in Fig.~\ref{fig:multiblockchain}(a) and (b), as the number of blockchain networks increases, the average throughput increases and the transaction fee rate decreases. The reason is that the increase of the number of blockchain networks allows the gateway to select a better blockchain network in terms of high security and low cost storage. As a result, the total reward increases as shown in Fig.~\ref{fig:multiblockchain}(c).

In summary, the simulation results shown in this section
confirm that the DRL algorithm is able to solve the computation expensive problem of the large action and state spaces of the Q-learning. Furthermore, the proposed DRL algorithm can be used for the gateway to learn the optimal policy. The optimal policy allows the gateway to optimally assign time slots to the secondary transmitters for the energy harvesting, data backscatter, and data transmission under the dynamics, uncertainty, and unpredictability of the blockchain-based RF-powered backscatter cognitive radio network. In addition, the optimal policy enables the gateway to decide the best blockchain network with the appropriate transaction fee rate. As a result, the optimal policy leads to the network throughput maximization and the transaction fee minimization.
\section{Conclusion}
In this paper, we have presented the DRL algorithm for the time scheduling, blockchain selection, and transaction fee rate decision in the blockchain-based RF-powered backscatter cognitive radio networks. Specifically, we have formulated the time scheduling, blockchain selection, and transaction fee rate decisions of the secondary gateway as a stochastic optimization problem. To solve the problem, we have developed a DRL algorithm using D3QN. The simulation results show that the proposed DRL algorithm enables the gateway to learn an optimal policy that maximizes the network throughput while minimizing the data storage cost. The throughput obtained by the proposed DRL algorithm is significantly higher than those of the baseline algorithms including Q-learning, random, HTT and backscatter policies.


\vspace{-0.3cm}
\begin{thebibliography}{00}

\bibitem{huynh2018}
N. V. Huynh, D. T. Hoang, D. N. Nguyen, E. Dutkiewicz, D. Niyato, and P. Wang, ``Reinforcement learning approach for RF-powered cognitive radio network with ambient backscatter,'' to be presented in {\em IEEE Global Communications Conference}, Abu Dhabi, UAE, 9-13 Dec. 2018.

\bibitem{li2018}
D.~Li, W.~Peng, and Y.-C.~Liang, ``Hybrid Ambient Backscatter Communication Systems with Harvest-Then-Transmit Protocols'', {\em IEEE Access}, vol. 6, no. 1, pp. 45288-45298, Dec. 2018.

\bibitem{kang2018}
X.~Kang, Y.-C.~Liang, and J.~Yang, ``Riding on the Primary: A New Spectrum Sharing Paradigm for Wireless-Powered IoT Devices'', {\em IEEE Transactions on Wireless Communications}, vol. 17, no. 9, pp. 6335-6347, Sept. 2018.

\bibitem{liu2013}
V. Liu, A. Parks, V. Talla, S. Gollakota, D. Wetherall, and J. R. Smith, ``Ambient backscatter: wireless communication out of thin air,'' in {\em ACM SIGCOMM Computer Communication Review}, vol. 43, no. 4, pp. 39-50, Aug. 2013.

\bibitem{khan2017cognitive}
A. A. Khan, M. H. Rehmani and A. Rachedi, "Cognitive-Radio-Based Internet of Things: Applications, Architectures, Spectrum Related Functionalities, and Future Research Directions," in {\em IEEE Wireless Communications,} vol. 24, no. 3, pp. 17-25, June 2017.

\bibitem{schrijvers2016}
Schrijvers, O. Bonneau, J. Boneh, D. Roughgarden, Tim, "Incentive compatibility of bitcoin mining pool reward functions," {\em International Conference on Financial Cryptography and Data Security,} Barbados, pp. 477--498, May 2016.

\bibitem{khan2018}
M. A. Khan, K. Salah, "IoT security: Review, blockchain solutions, and open challenges, '' in {\em Future Generation Computer Systems,} vol. 82, no. 1, pp. 395--411, 2018.

\bibitem{kim2019blockchained}
H. Kim, J. Park, M. Bennis, and S. L. Kim, "Blockchained on-device federated learning, '' in {\em IEEE Communications Letters,} early access.


\bibitem{neisse2017}
R. Neisse, G. Steri, and I. Nai-Fovino, “A blockchainbased approach for data accountability and provenance tracking,” in {\em Proceedings of the 12th International Conference on Availability, Reliability and Security,} Reggio Calabria, Italy: ACM, pp. 93--98, August 2017.

\bibitem{liu2017}
B. Liu, X. L. Yu, S. Chen, X. Xu, and L. Zhu, “Blockchain based data integrity service framework for iot data,” in {\em IEEE International Conference on Web Services (ICWS),} Honolulu, HI, pp. 468--475, June 2017.

\bibitem{saghiri2018}
A. M. Saghiri, M. Vahdati, K. Gholizadeh, M. R. Meybodi, M. Dehghan, and H. Rashidi, ``A framework for cognitive Internet of Things based on blockchain,'' in {\em IEEE International Conference on Web Research (ICWR),} Tehran, Iran, pp. 138-143, Apr. 2018.

\bibitem{mnih2015}
V. Mnih, K. Kavukcuoglu, D. Silver, A. A. Rusu, J. Veness, M. G. Bellemare, A. Graves, M. Riedmiller, A. K. Fidjeland, G. Ostrovski {\em et al.,} ``Human-level control through deep reinforcement learning,''
Nature, vol. 518, no. 7540, p. 529--533, Feb. 2015.


\bibitem{luong2019surveydrl}
N.~C.~Luong, D.~T.~Hoang, S. Gong, D.~Niyato, P.~Wang, Y. C. Liang, and D. I. Kim, ``Applications of Deep Reinforcement Learning in Communications and Networking: A Survey,'' {\em  IEEE Communications Surveys and Tutorials}, to appear.



\bibitem{hoang2017}
D.~T.~Hoang, D.~Niyato, P.~Wang, D.~I.~Kim, and L.~Bao Le, ``Optimal Data Scheduling and Admission Control for Backscatter Sensor Networks,'' {\em IEEE Transactions on Communications}, vol. 65, no. 5, pp. 2062--2077, May 2017.


\bibitem{lyu2018a}
B. Lyu, C. You, Z. Yang, and G. Gui, ``The Optimal Control Policy for RF-Powered Backscatter Communication Networks,'' {\em IEEE Transactions on Vehicular Technology}, vol. 67, no. 3, pp. 2804--2808, Mar. 2018.

\bibitem{lyu2018b}
B. Lyu, H. Guo, Z. Yang, and G. Gui, ``Throughput Maximization for Hybrid Backscatter Assisted Cognitive Wireless Powered Radio Networks,'' {\em IEEE Internet of Things Journal}, vol. 5, no. 3, pp. 2015--2024, Jun. 2018.

\bibitem{yang2018}
Q. Yang, H.-M. Wang, T.-X. Zheng, Z. Han, and M. H. Lee, ``Wireless Powered Asynchronous Backscatter Networks With Sporadic Short Packets: Performance Analysis and Optimization,'' {\em IEEE Internet of Things Journal}, vol. 5, no. 2, pp. 984--997, Apr. 2018.

\bibitem{kwan2018}
J. C. Kwan and A. O. Fapojuwo, ``Sum-Throughput Maximization in Wireless Sensor Networks With Radio Frequency Energy Harvesting and Backscatter Communication,'' {\em IEEE Sensors Journal}, vol. 18, no. 17, pp. 7325--7339, Sept. 2018.


\bibitem{gong2018}
W. Gong, H. Liu, J. Liu, X. Fan, K. Liu, Q. Ma, and X. Ji, ``Channel-Aware Rate Adaptation for Backscatter Networks,'' {\em IEEE/ACM Transactions on Networking}, vol. 26, no. 2, pp. 751--764, Apr. 2018.

\bibitem{wang2018}
W. Wang, D. T. Hoang, D. Niyato, P. Wang, and D. I. Kim, ``Stackelberg game for distributed time scheduling in RF-powered backscatter cognitive radio networks'', in {\em IEEE Transactions on Wireless Communications}, vol. 17, no. 8, pp. 5606--5622, Jun. 2018. 

\bibitem{gao2019}
X. Gao, P. Wang, D. Niyato, K. Yang, and J. An, “Auction-based time scheduling for backscatter-aided RF-powered cognitive radio network,” {\em IEEE Transactions on Wireless Communications,} vol. 18, no. 3, pp. 1684--1697, Mar. 2019.

\bibitem{anh2019}
T. T. Anh, N. C. Luong, D. Niyato, Y. C. Liang, and D. I. Kim, "Deep Reinforcement Learning for Time Scheduling in RF-Powered Backscatter Cognitive Radio Networks," {\em IEEE Wireless Communications and Networking Conference (WCNC)}, April 2019, Marrakech, Morocco.

\bibitem{kotobi2017}
K. Kotobi and S. G. Bilén, ``Blockchain-enabled spectrum access in cognitive radio networks,`` {\em 2017 Wireless Telecommunications Symposium (WTS)}, Chicago, IL, pp. 1--6, 2017.

\bibitem{raju2017}
S. Raju, S. Boddepalli, S. Gampa, Q. Yan and J. S. Deogun, "Identity management using blockchain for cognitive cellular networks," {\em 2017 IEEE International Conference on Communications (ICC)}, Paris,  pp. 1-6, 2017.

\bibitem{shaohan2019}
S. Feng, W. Wang, D. Niyato, D. I. Kim, and P. Wang, “Crowdsensing Market for Wireless-Powered Internet of Things with Scalable Blockchains,” {\em IEEE Transactions on Wireless Communications,} submitted.

\bibitem{Cauchytheorem}
E. of Mathematics, ``Cauchy-lipschitz theorem,'' https://www.en
cyclopediaofmath.org/index.php/Cauchy-Lipschitz theorem.

\bibitem{weber2017availability}
I. Weber, V. Gramoli, A. Ponomarev, M. Staples, R. Holz, A. B. Tran, and P. Rimba, ``On availability for blockchain-based systems.'' {\em IEEE 36th Symposium on Reliable Distributed Systems (SRDS)}, pp. 64--73, Sept. 2017.

\bibitem{nguyen2019}
N. C. Luong, T.T. Anh, H.T.T. Binh, D. Niyato, D.I. Kim, Y. Liang, ``Joint Transaction Transmission and Channel Selection in Cognitive Radio Based Blockchain Networks: A Deep Reinforcement Learning Approach.`` {\em IEEE International Conference on Acoustics, Speech and Signal Processing (ICASSP)}, pp. 8409--8413, 2019.


\bibitem{wang2016}
Wang Z, Schaul T, Hessel M, Van Hasselt H, Lanctot M, De Freitas N., ``Dueling network architectures for deep reinforcement learning.`` arXiv preprint arXiv:1511.06581. 2015 Nov 20.

\bibitem{kumar2019proof}
G.~Kumar, R.~Saha, M.~K. Rai, R.~Thomas, and T.-H. Kim, ``Proof-of-work
  consensus approach in blockchain technology for cloud and fog computing using
  maximization-factorization statistics,'' \emph{IEEE Internet of Things
  Journal}, to appear.
  
  \bibitem{yang2018blockchain}
  Z. Yang, K. Yang, L. Lei, K. Zheng, and V. C. Leung, ``Blockchain-based decentralized trust management in vehicular networks,'' \emph{IEEE Internet of Things
  Journal}, vol. 6, no. 2, pp. 1495-1505, Apr. 2018.
 

\bibitem{transaction_fee}
(2018, Sept.) Explaining bitcoin transaction fees. [Online]. Available: https://support.blockchain.com/hc/en-us/articles/


\bibitem{karame2012double}
G.~O. Karame, E.~Androulaki, and S.~Capkun, ``Double-spending fast payments in
  bitcoin,'' in \emph{ACM conference on Computer and communications security},
  Raleigh, NC, October 2012, pp. 906--917.
  
  \bibitem{rosenfeld2014analysis}
M.~Rosenfeld, ``Analysis of hashrate-based double spending,'' \emph{arXiv
  preprint arXiv:1402.2009}, 2014.
  
  
\bibitem{watkins1992q}
C.~J. Watkins and P.~Dayan, ``Q-learning,'' \emph{Machine learning}, vol.~8,
  no. 3-4, pp. 279--292, May 1992.

\bibitem{consul1973}
P. C. Consul and G. C. Jain, ``A Generalization of the Poisson Distribution,`` {\em Technometrics}, vol. 15, no. 4, 791-799, Nov. 1973.


\bibitem{chen2017}
M. Chen, W. Saad, and C. Yin, ``Liquid state machine learning for resource allocation in a network of cache-enabled lte-u uavs,'' in {\em IEEE Global Communications Conference}, Singapore, pp. 1--6, Dec. 2017.

\bibitem{wang2017}
S. Wang, H. Liu, P. H. Gomes, and B. Krishnamachari, ``Deep reinforcement learning for dynamic multichannel access in wireless networks,'' in {\em IEEE Transactions on Cognitive Communications and Networking}, vol. 4, no. 2, pp. 257--265, Feb. 2018.



\bibitem{he2017}
Y. He, Z. Zhang, and Y. Zhang, ``A big data deep reinforcement learning
approach to next generation green wireless networks,'' in {\em IEEE Global Communications Conference}, Singapore, pp. 1--6, Dec. 2017.

\bibitem{google2017}
B.~McMahan and D.~Ramage, ``Federated learning: Collaborative machine learning
without centralized training data,'' https://ai.googleblog.com/2017/
04/federated-learning-collaborative.html, Accessed April 6, 2017.

\bibitem{shehabi2018}
Al-Shehabi, Abdullah, ``Bitcoin Transaction Fee Estimation Using Mempool State and Linear Perceptron Machine Learning Algorithm``, Master's Projects, 2018.

\end{thebibliography}

\end{document}